\pdfoutput=1 
\documentclass[prr,twocolumn,superscriptaddress,longbibliography,nofootinbib]{revtex4-1}
\usepackage{latexsym}
\usepackage{lineno}
\usepackage{hyperref}
\usepackage{url}
\usepackage[caption=false,font={normalsize}]{subfig}
\usepackage{graphicx}
\usepackage{verbatim}
\usepackage{multirow}
\usepackage{amsmath, bm}
\newcommand{\beq}{\begin{eqnarray}}
\newcommand{\eeq}{\end{eqnarray}}
\usepackage{mathrsfs}
\usepackage{float}
\usepackage{bbm}

\usepackage{mathtools}
\usepackage{slashed}
\usepackage{physics}	% installed packages for typing maths and physics
\usepackage{graphicx}   % need for Fig.ures
\usepackage{epstopdf}
\usepackage{hyperref}   % use for hypertext links, including those to external documents and URLs
\usepackage{bbold}
\usepackage{wasysym}
\usepackage{amssymb}
\usepackage{soul}
\usepackage[symbol]{footmisc}
\def \bs{\textbf}
\usepackage{tikz}
\usetikzlibrary{decorations.pathmorphing}
\usetikzlibrary{shapes.misc}
\tikzset{cross/.style={cross out, draw=black, minimum size=8*(#1-\pgflinewidth), inner sep=0pt, outer sep=0pt},
cross/.default={1pt}}
\usetikzlibrary{patterns,math}
\newcommand{\RN}[1]{%
  \textup{\uppercase\expandafter{\romannumeral#1}}%
}

\begin{document}
\title{Cubic Dirac Semimetals: General Theory and 
Application to Rare-Earth Magnets
}
\author{Shouvik Sur}
\affiliation{Department of Physics and Astronomy, Rice Center for Quantum Materials, Rice University, Houston, Texas 77005, USA}
\author{Chandan Setty$^{\dagger}$}
\affiliation{Department of Physics and Astronomy, Iowa State University, Ames, Iowa 50011, USA}
\affiliation{Ames National Laboratory, U.S. Department of Energy, Ames, Iowa 50011, USA}

\begin{abstract} 
Rare-earth magnets with parent cubic symmetry exhibit unique topological properties. 
However, the origin of 
these behaviors remains presently unclear. Here, we develop minimal models for Dirac semimetals (DSMs) with accidental band crossings and higher-order topology in cubic systems, incorporating candidate magnetic order 
to analyze bulk, surface, and hinge state characteristics. In certain cubic-symmetric DSMs, we identify an effective  $\mathbb{Z}_2$  chiral symmetry which significantly impacts surface and hinge-localized states. Our results highlight distinct features in surface state dispersions, Fermi arcs, polarization dependence, and band splitting that correlate with photoemission data in rare-earth monopnictides. We also suggest candidate materials and experimental tests for further validation. These findings advance our understanding of surface states in rare-earth magnets with parent cubic symmetries and illuminate the role of DSM physics in these systems.
\end{abstract}

\maketitle
\section{Introduction}
Magnetic topological semimetals with cubic symmetries in the parent state, particularly those found in rare-earth magnets with rock-salt structures~\cite{Ong2016, CK2022-Nat, CK2023-CP, CK2022-PRB, CK2023-PRB, CK-PRB2019}, have recently gained attention. 
The cubic space group permits several point group symmetries that can protect topological band crossings~\cite{ryu2016}.
These include three- and four-fold rotational symmetries which
can individually  stabilize crossings between pairs of Kramer's degenerate bands, giving rise to three-dimensional Dirac semimetals  (DSMs)~\cite{nagaosa2014}.
Therefore, cubic crystals offer a promising platform for realizing cubic-symmetric DSMs.

Over the past decade, several cubic-symmetric materials have been proposed as  DSM candidates. 
Notably,  \emph{ab initio} calculations have identified several full  Heusler 
alloys~\cite{guo2017type, mondal2019type}, along with  Pd and Pt oxides as potential DSMs~\cite{li2017topological}.
More recently, experiments on the cubic material PtBi$_2$~\cite{CK-PRB2019} have revealed existence of Fermi-arc like surface states, leading to the suggestion that it is an DSM.
Moreover, cubic-symmetric DSM  states have been proposed as parent phases for magnetic Weyl semimetals~\cite{Cao2018, Yuan2017, Zhu2020}.
Despite the growing body of literature supporting the existence of DSMs in cubic crystals, the extent to which Dirac semimetallic features influence the electronic behavior of cubic crystalline materials remains unclear and has yet to be systematically explored.
%%

%%%%%%%%%%%%%%%%%%%%%%%%%%%
\begin{figure}[!t]
\centering
\subfloat[]{%
\includegraphics[width=0.75\columnwidth]{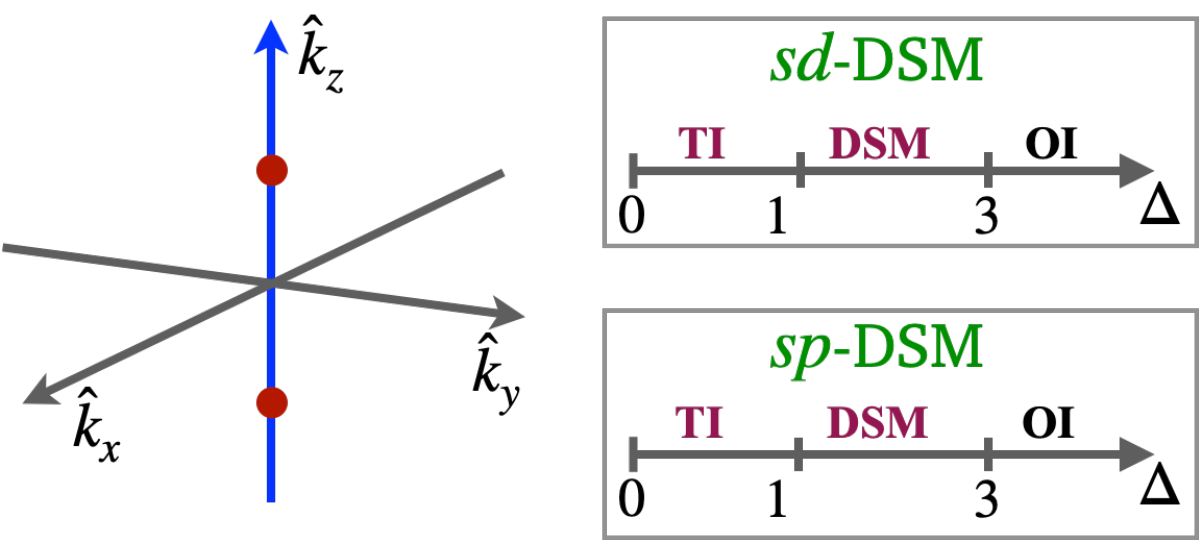}%
}
\hfill
\subfloat[]{%
\includegraphics[width=0.75\columnwidth]{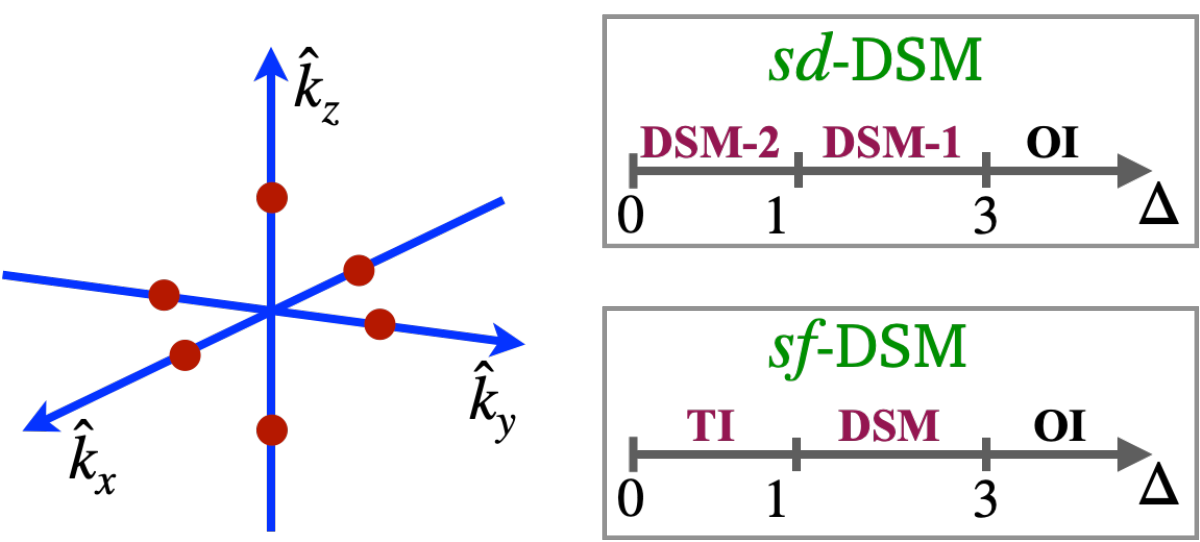}%
}
\caption{Universal bulk-characteristics and phase diagram of symmetry-protected Dirac semimetals (DSMs). 
(a) The minimum number of Dirac points (red dots; DPs) in rotational-symmetry protected DSMs is 2, which lie along the rotation axis (highlighted in blue). 
(b) In cubic-symmetry protected DSMs the minimum number of DPs is 6 which appear along the three symmetry-equivalent, fourfold-rotation axes.
In inversion and time-reversal symmetry preserving systems, there are two types of DSMs, $sd$ and $sp$ [$sd$ and $sf$], in category (a) [(b)].
(inset) Respective phase diagrams supported by the minimal Hamiltonians describing these phases on   (a) tetragonal  and (b) simple cubic lattices. 
Here, ``TI'' and ``OI'' are acronyms for topological insulator and ordinary insulator.
}
\label{fig:DSMs}
\end{figure}
%%%%%%%%%%%%%%%%%%%%%%%%%%%

Understanding cubic-symmetric DSMs poses two key challenges, however.
First, the multiple equivalent rotation axes and mirror planes in cubic systems can protect a variety of bulk-band crossings, complicating the interpretation of experimental data. 
As a result, early theoretical works on DSMs primarily focused on trigonal, tetragonal, and hexagonal systems, which feature a single principal rotation  axis. 
Through minimal Hamiltonians, various aspects of the physics of rotational-symmetry protected DSMs have been addressed, including   nature of band-topology~\cite{lu2016, Randeria2018, Bernevig2020, Goswami2023, Cano2021}, classification schemes~\cite{nagaosa2014, furusaki2015,  zhang2016}, and experimental implications~\cite{kim2016, AMV-RMP}. 
Insights from such minimal modeling have led to the identification of several candidate materials~\cite{fang2012, fang2013, Hu2017, Hu2018, Kim2019, zhang2016a}, and interpretation of experimental investigations into  tetragonal~\cite{YLChen2014, Hasan2015-3, Cava2014, Yaresko2020, ShikShin2019, Ding2015, Ding2021} and hexagonal~\cite{YLChen2014-2, Hasan2015-4, Ong2015, Zhou2017, Hasan2023} symmetric Dirac materials.

Second, theoretical investigations have revealed an absence of a robust bulk-boundary correspondence in rotational-symmetry protected DSMs~\cite{lu2016, Randeria2018, Bernevig2020, Goswami2023, Cano2021}, in sharp contrast to topological insulators and Weyl semimetals~\cite{note1}. 
Since surface states have been pivotal in probing the bulk topology of topological materials~\cite{AMV-RMP, Ding2021-RMP}, this feature of symmetry-protected DSMs complicates the diagnosis of their bulk-topology using  surface-sensitive techniques  such as angle resolved photoemission spectroscopy (ARPES) and scanning tunneling microscopy (STM).  
Some of this ambiguity can yet be resolved by examining  the universal features of minimal models, which offer a clearer framework for  associating surface state characteristics with bulk topological properties.

Here, we systematically develop a theory of non-magnetic DSMs in cubic crystals, and demonstrate the universal aspects of the  relationship between various space group symmetries and observable properties of such DSMs.
We find two classes of cubic-symmetric DSMs, based on the parity of the orbitals that hybridize to give rise to  Dirac points (DPs).
In analogy to similar classification of rotational-symmetry protected DSMs~\cite{Bernevig2020}, we name them $sd$-DSM (orbitals with same parity) and $sf$-DSM (orbitals with opposite parity).
In both cases the minimum number of DPs is six, which is three times larger than those in rotational-symmetry protected DSMs; see Fig.~\ref{fig:DSMs}.
This feature indicates that the surface states in cubic-symmetric DSMs are qualitatively distinct than those in rotational-symmetry protected DSMs. 
We construct minimal models to study the topology of the Bloch wavefunctions, identify the universal symmetry-protected features of the surface states, and the nature of spin-momentum locking the latter supports. More specifically, in the $sd$-DSM we identify an effective  $\mathbb{Z}_2$  chiral symmetry which significantly impacts surface and hinge-localized states.
Further, by incorporating candidate magnetic orders that may develop as instabilities of  cubic-symmetric DSMs, we show that this framework could provide insights into key spectral features observed in rare-earth magnets. 
We further make specific predictions that can validate the role of DSM physics and also identify potential cubic materials that could be experimentally tested for similar phenomena.
Overall, our study suggests that DSM physics may offer a coherent framework for understanding surface states in magnetic topological semimetals with cubic parent symmetries, connecting various materials, and clarifying their origins.

\section{Experiments in REPns} 
Recent photoemission studies~\cite{CK2022-Nat, CK2023-CP, CK2022-PRB, CK2023-PRB} on rare-earth monopnictides (REPns) with a rock-salt crystal structure have revealed intriguing surface state features the origins of which remain poorly understood. These include unusual surface state splittings (\textcolor{black}{Kaminski-Canfield splittings~\cite{CK2022-Nat, CK-2024CommMat, CK-2024PRB}}), unique Fermi arcs and pockets as well as polarization dependencies~\cite{CK2022-Nat}.  In their high-temperature phase, REPns adopt a paramagnetic face-centered cubic lattice (SG 225). Density Functional Theory (DFT) calculations indicate that the bulk electronic structure shows Dirac-like band crossings away from the Fermi energy and topological insulator (TI)-like anti-crossings closer to it~\cite{CK2023-CP}. Magnetization and magnetic susceptibility measurements across several REPn compounds suggest a zero-field antiferromagnetic (AFM) transition~\cite{CK2022-Nat, CK2022-PRB, CK2023-PRB}. While complex two- and three-dimensional AFM spin ordering patterns have been proposed~\cite{CK2023-CP}, the precise magnetic space group and the exact nature of the magnetic ordering remain unclear, leaving the location of DPs in the magnetic phase similarly uncertain.
The degree of electron correlations also varies among REPn family members, ranging from moderate to strong, further complicating the identification of DPs. Thus, it remains unclear whether the observed surface state features originate from a TI, bulk DPs, or features induced by magnetic order, either in the TI phase or in a DSM. 
This ambiguity, along with experimental uncertainties and the material's complex properties, makes theoretical investigations particularly challenging. However, as we discuss below, our minimal models, despite their simplicity, can shed light on these features and serve as a critical foundation for developing even more material-specific models in the future. 

%%%%%%%%%%%%%%%%%%%%%%%%%%%%%%%%%%%%
\begin{figure}[!t]
    \centering
    \includegraphics[width=8.0cm, height=4.25cm]{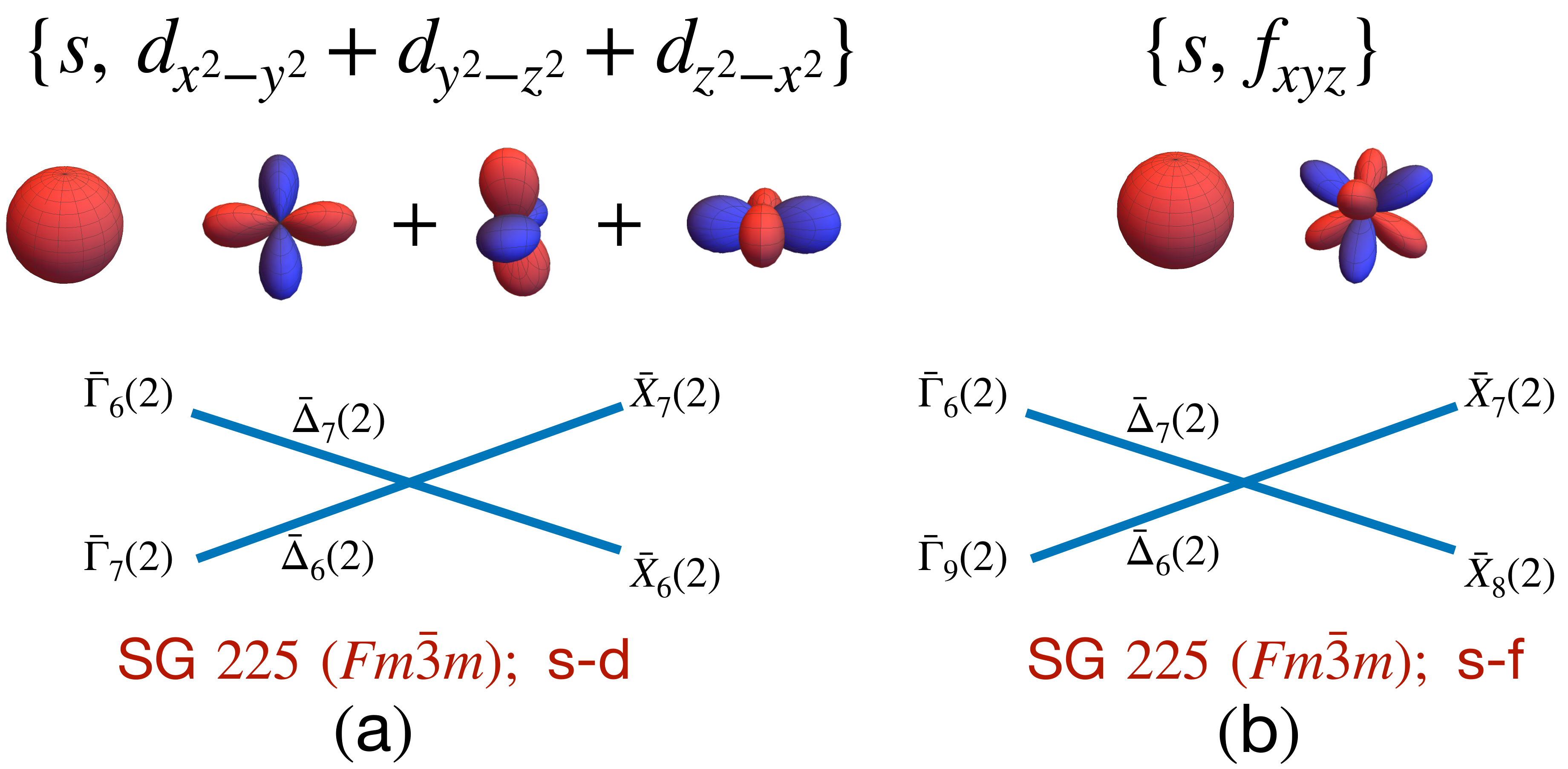}
    \caption{Schematic figure showing the $s$-$d$ and $s$-$f$ orbitals and representations along high-symmetry points and lines}
    \label{Fig1}
\end{figure}
%%%%%%%%%%%%%%%%%%%%%%%%%%%%%%%%%%%%

\section{Models} \label{sec:models}
We begin by recalling the generators of the cubic space group. The fundamental symmetry operations are two-fold rotations around the $z$-axis, $\hat{C}_{2z} = \{2_{001} | 0\}$, the $y$-axis, $\hat{C}_{2y} = \{2_{010} | 0\}$, the diagonal axes $\hat{C}_{2xy} = \{2_{110} | 0\}$, the three-fold rotation around the $[111]$ axis, $\hat{C}_3 = \{3^{+}_{111} | 0\}$ and inversion symmetry $\hat{I} = \{-1 | 0\}$. In addition, we also impose time-reversal symmetry $\hat{T}$ that ensures Kramer's degeneracy throughout the Brillouin zone. 
Four-fold rotations can be constructed as combinations of two-fold rotations in the horizontal and diagonal planes.
We focus on the simplest realization of four-fold rotational-symmetry-protected DSMs in cubic systems.  We limit our scope to four-band double representations for SG 225 ($Fm\bar 3m$), the space group relevant for the REPns with two orbitals and spins each. The four-fold rotational axes occur along the three equivalent $\Gamma-X$ (or $\Delta$) directions. The little group representations of interest transform as $\bar{\Delta}_6(2)$ and $\bar{\Delta}_7(2)$ with distinct rotational eigenvalues of $e^{\pm i \frac{3 \pi}{4}}$ and $e^{\pm i\frac{\pi}{4}}$ respectively that protect the topological crossings. These representations can emerge in two distinct ways depending on the orbital content -- whether the relative parity between the two orbitals is even or odd. A schematic of the two possibilites appears in Fig.~\ref{Fig1}. For the even relative parity combination, the simplest non-trivial scenario involves the $s$ and $d$ orbitals~\cite{nagaosa2014, Bernevig2020}. Although this can generally lead to a 12-band representation due to the involvement of all five $d$ orbitals, these bands can generally be decomposed into smaller representations. 
To replicate the correct transformation properties within a two-orbital ``$s$-$d$" model, we choose an effective orbital representation. One realization of the even parity orbital combination involves the local orbitals $|s\rangle$ and a combination of $d$ orbitals, $|d_{x^2 - y^2} + d_{y^2 - z^2} + d_{z^2 - x^2}\rangle$ appearing in Fig.~\ref{Fig1}(a). This draws from the analogy of $s$-$d$ DSMs in tetragonal and hexagonal systems~\cite{Bernevig2020, Goswami2023}.  

For the odd combination of relative parity, a natural combination of orbitals is $s$ and $p$ orbitals. However, in two-orbital double representations of the cubic symmetries, the additional mirror reflections and $C_3$ rotations preclude this scenario.  Higher dimensional representations involving $s$-$p$ orbitals may however be possible. This must be contrasted with hexagonal and tetragonal DSMs where two orbital double representations involving $s$-$p$ orbitals are allowed. The simplest possible odd (relative) parity four-band representation involves the local orbitals $|s\rangle$ and $|f_{xyz}\rangle$ (see Fig.~\ref{Fig1}(b)). We henceforth refer to this as the ``$s$-$f$" scenario.

Except for the inversion operator, the matrix representations of the cubic generators take the same form for the $s$-$d$ and $s$-$f$ orbitals. They are given by $\hat{C}_{2z} = \tau_0 \otimes \exp[-i\frac{\pi}{2} \sigma_3]$, $\hat{C}_{2y} = \tau_0 \otimes i \sigma_2$, $\hat{C}_{2xy} = \tau_3 \otimes \exp[-i\frac{\pi}{2\sqrt{2}} (\sigma_1 + \sigma_2)]$, $\hat{C}_3 = \tau_0 \otimes \exp[i\frac{\pi}{3\sqrt{3}} (\sigma_1 + \sigma_2 + \sigma_3)]$ while the time-reversal operator for spinful systems is $\hat{T} = (\tau_0 \otimes -i \sigma_2) \hat{K}$,
where $\hat{K}$ is the complex conjugation operator. The inversion operator for the $s$-$d$ ($s$-$f$) orbitals is $\hat{I} = \tau_0 \otimes \sigma_0 (\tau_3 \otimes \sigma_0)$. 
Here, $\tau_j$/$\sigma_j$ [$\tau_0$/$\sigma_0$] is the $j$-th Pauli [$2\times 2$ identity matrix] matrix with   $\tau_\mu (\sigma_\mu)$ acting on the orbital (spin) degree of freedom.

The lattice Hamiltonian in the presence of time reversal symmetry is 
\begin{align}
H_0^{d,f} &= -\sum_{\substack{ij \alpha \beta \\ \sigma \sigma'}} t_{ij \sigma \sigma'}^{\alpha\beta} c_{i\alpha \sigma}^\dagger c_{j\beta \sigma'} + h.c. - \mu \sum_{i \alpha \sigma} c_{i\alpha \sigma}^\dagger c_{i\alpha \sigma} \\
&= \sum_k \Psi_k^\dagger H_0^{d,f}(k_x, k_y, k_z) \Psi_k,
\end{align}
where $t_{ij\sigma \sigma'}^{\alpha\beta}$ are the hopping parameters between site $i$ and $j$, orbitals $\alpha$ and $\beta$, and spins $\sigma$ and $\sigma'$.  $c_{i\alpha \sigma}^\dagger$ denotes the electron creation operator for site $i$, orbital $\alpha$ and spin $\sigma$, $\mu$ the chemical potential and we work in the basis $\Psi_k^\dagger = \left( c_{\bs k1\uparrow}^{\dagger}, c_{\bs k1\downarrow}^{\dagger}, c_{\bs k2\uparrow}^{\dagger}, c_{\bs k2\downarrow}^{\dagger} \right)$. The index $``1"$ indicates the orbital $s$ and $``2"$ labels the $d$ and $f$ orbital for $H_0^d$ and $H_0^f$ respectively.  
\textcolor{black}{The corresponding Bloch  Hamiltonians are obtained by keeping all terms that are allowed by the cubic generators for $s$-$d$ and $s$-$f$ orbitals. This leads to the Hamiltonian matrices given by}
\begin{align}  \nonumber
H_0^d(\mathbf{k}) &= t_1^d \left(s_{k_x} s_{k_z} \Gamma_{15} + s_{k_y} s_{k_z} \Gamma_{25} + s_{k_x} s_{k_y} \Gamma_{35}\right) \\ \nonumber
&+ t_2^d \left( \sum_{j=x,y,z} c_{k_j} - \Delta \right) \Gamma_5  \\ \nonumber
&+\qty[w_1^d \sum_{j=x,y,z} c_{k_j} 
+ w_2^d \prod_{j=x,y,z} c_{k_j} - \mu] \mathbbm{1} \\ \nonumber
H_0^f(\mathbf{k}) &= t_1^f \Big(s_{k_x} (c_{k_z} - c_{k_y})  \Gamma_1 + s_{k_y} (c_{k_x} - c_{k_z})  \Gamma_2 \\ \nonumber
&+  s_{k_z} (c_{k_y} - c_{k_x}) \Gamma_3 \Big) 
+ t_3^f s_{k_x}s_{k_y} s_{k_z} \Gamma_4 \\ \nonumber
&+ t_2^f \left(\sum_{j=x,y,z} c_{k_j} - \Delta \right) \Gamma_5 \\
&+\qty[w_1^f \sum_{j=x,y,z} c_{k_j} 
+ w_2^f \prod_{j=x,y,z} c_{k_j} - \mu ] \mathbbm{1}.
\label{Eq:HdHf}
\end{align}
Here, the $\Gamma$ matrices are defined as:
\[
\Gamma_{j=1,2,3} = \tau_1 \otimes \sigma_j, \quad \Gamma_4 = \tau_2 \otimes \sigma_0, \quad \Gamma_5 = \tau_3 \otimes \sigma_0,
\]
with the anticommutation relation $\{\Gamma_i, \Gamma_j\} = 2 \delta_{ij}$. 
In addition to $\Gamma_i$,  the Hamiltonian and symmetry operators can  be written  in terms of $\Gamma_{ij} = \frac{1}{2i}[\Gamma_i, \Gamma_j]$ for $i < j = 1, \dots, 5$ since $\{ \Gamma_i, \Gamma_{ij} \}$ together form the fifteen generators of SU(4) algebra. 
The notation $s_{k_i}, c_{k_i}$ is a short hand for sines and cosines of $k_i$. 
Note that we have used the $T_{2g} \left[T_{2u}/A_{2u}\right]$ harmonics of the $O_h$ point group to construct the off-diagonal terms for $H_0^d(\mathbf{k}) \left[H_0^f(\mathbf{k})\right]$. 
Therefore, the lowest-order off-diagonal terms in $H_0^d(\mathbf{k}) \left[H_0^f(\mathbf{k})\right]$ is quadratic (third order).  
The four-dimensional unit matrix is denoted by $\mathbbm{1}$ contains terms allowed by symmetry. 
Finally, $t^{d,f}_{1,2}, t_3^{f}, w^{d,f}_{1,2}$ are the respective hopping parameters of the tight binding model for $H_0^d(\mathbf{k}), H_0^f(\mathbf{k})$, and $\Delta$ is the mass parameter that tunes between different topological phases. 
We note that the combination of the space group symmetries and the relative parity between the hybridizing orbitals suppresses the $\Gamma_4$ matrix in the $H_0^d(\bs k)$, which results in a $\mathbb{Z}_2$ chiral symmetry for the four-band model of $sd$-DSMs in the limit where both $w_n^d = 0$ and we obtain  $\{H_0, \Gamma_4\} = 0$.

\begin{figure*}[!t]
\centering
\subfloat[]{%
\includegraphics[width=0.35\linewidth]{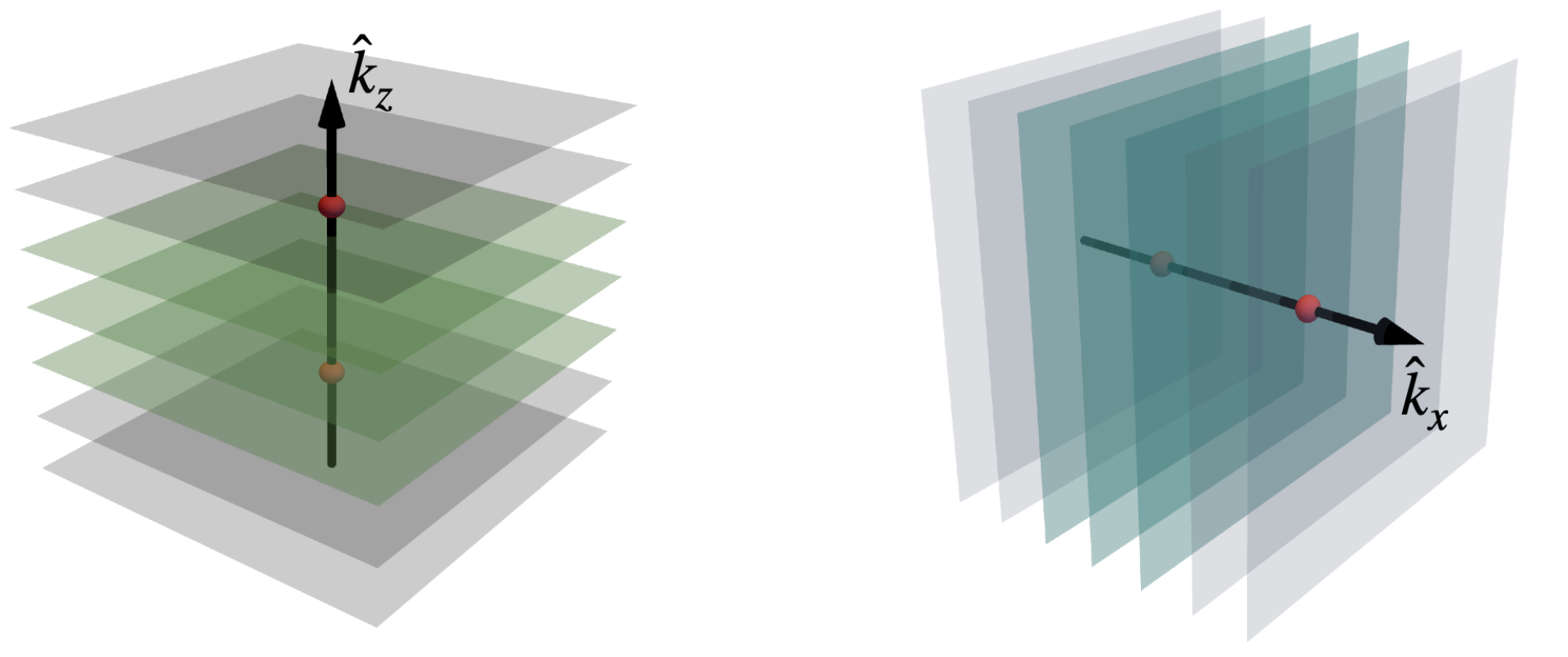}%
}
\hfill
\subfloat[]{%
\includegraphics[width=0.4\linewidth]{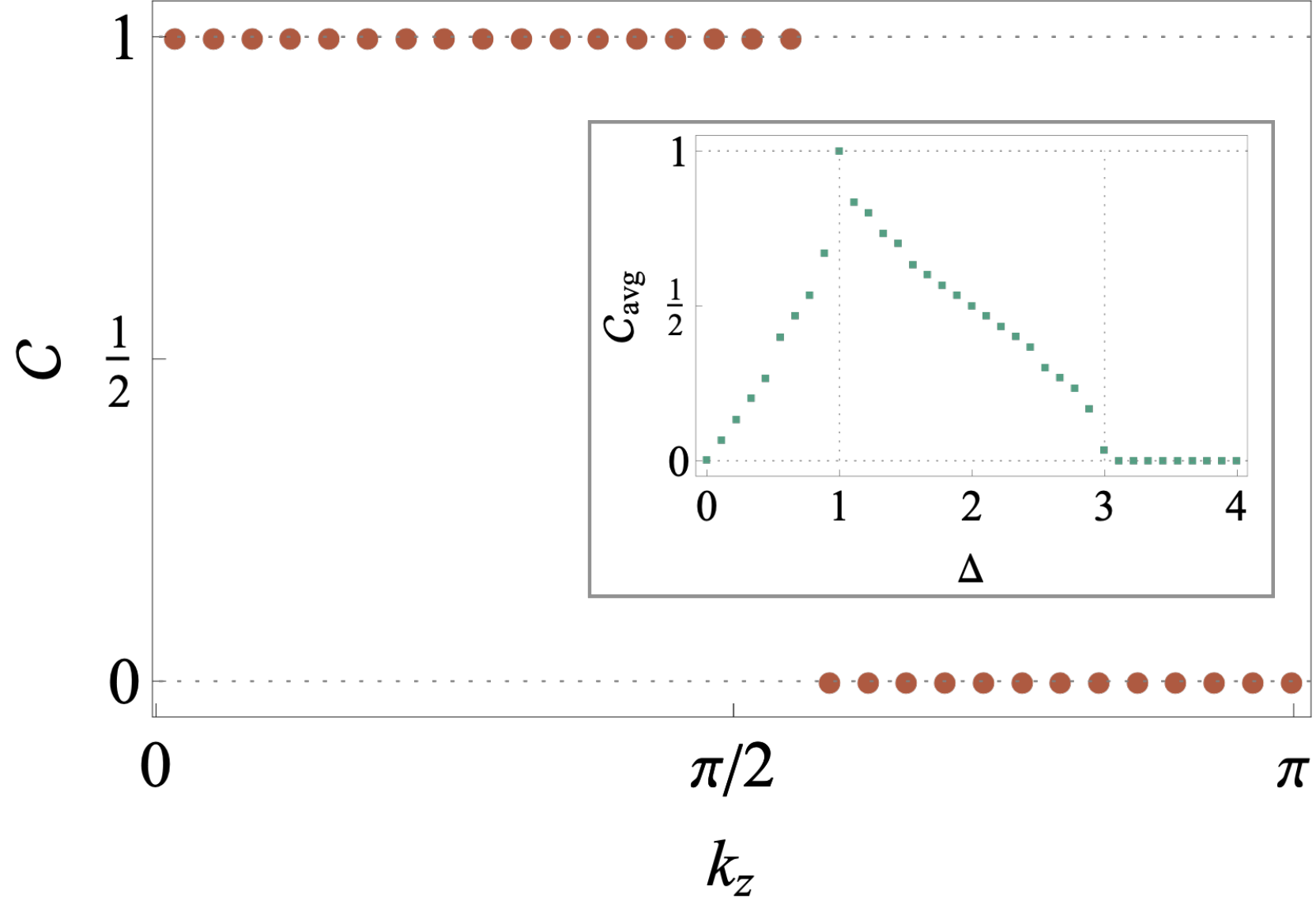}%
}
\hfill
\subfloat[]{%
\includegraphics[width=0.25\linewidth]{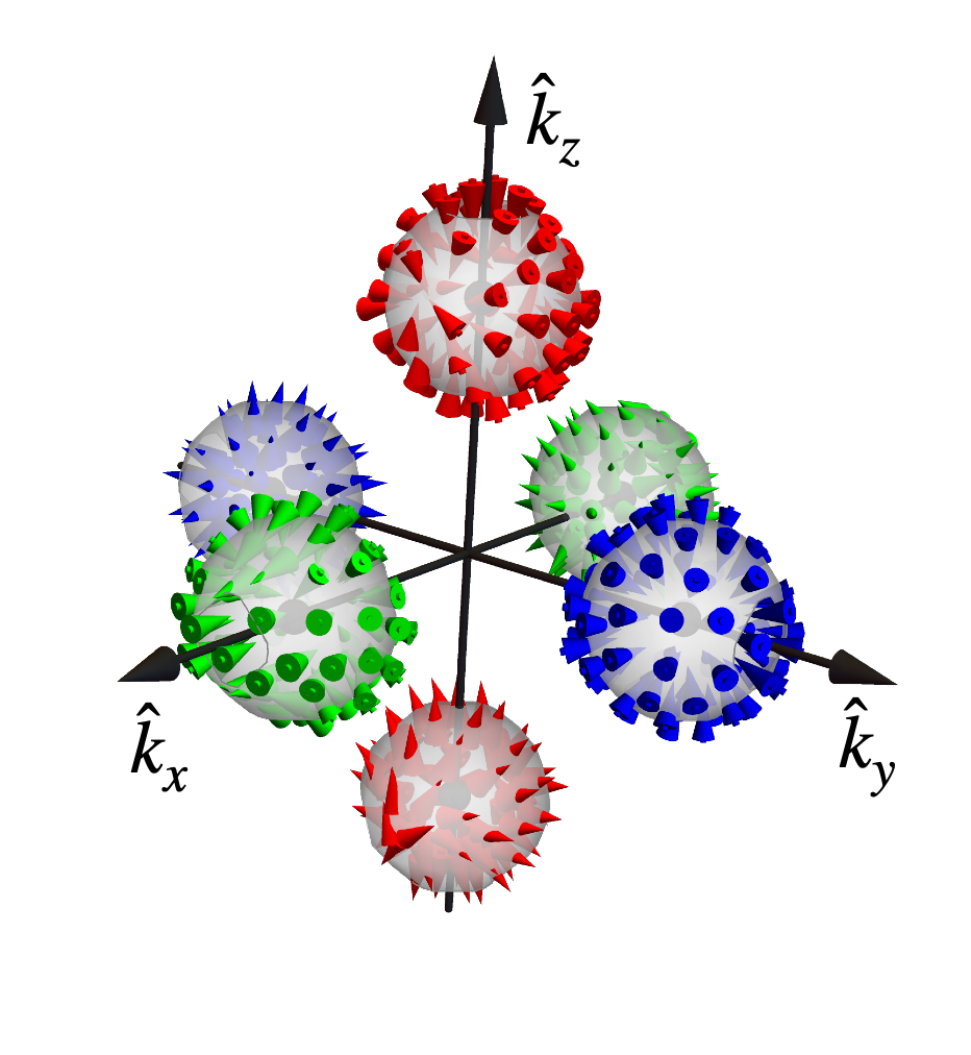}%
}
\caption{Bulk topology of cubic-symmetric Dirac semimetals (DSMs). (a) The planes (colored) separating a pair of Dirac points (DPs) along a rotation axis (red dots) support a quantized Berry flux [see (b)], while planes outside this region are topologically trivial (gray). 
Three independent sets of such planes exist, each set stacked along a rotation axis that supports DPs [cf. Fig~\ref{fig:DSMs}]. 
Here, we show only $\hat k_x$ and $\hat k_y$ axes for convenience. 
(b) The relative-Chern number $(\mathcal C)$ supported by $k_x-k_y$ planes  in the DSM-1 phase of the $sd$-DSM as a function of $k_z$. 
It jumps across the DP carrying plane. 
(inset) Average relative-Chern number ($\mathcal C_{\text{avg}}$), as defined in the main text, as a function of the band-inversion parameter $\Delta$. 
$\mathcal C_{\text{avg}}$ reflects the phase diagram in Fig.~\ref{fig:DSMs}(b) for $sd$-DSM.
(c) Depiction of the texture of the projected Berry curvature in the vicinity of the six DPs. 
The  green, blue, and red colors indicate projections on the fourfold rotation generators about the $\hat x$, $\hat y$, and $\hat z$ axes, respectively. 
\textcolor{black}{For both figures in (b) we have used $(t_1^d, t_2^d, w_j^d, \mu) = (1.4, 1.0, 0.0, 0.0)$, while for the main figure we have also fixed $\Delta = 1.8$.}
}
\label{fig:bulk-topo}
\end{figure*}

\subsection{Topology and phase diagram}
In rotation-symmetry protected DSMs, the planes in the Brillouin zone that are perpendicular to the principal rotation-axis can  be classified as two-dimensional higher-order topological insulators (HOTIs), with the DPs lying at the boundary between topologically trivial and non-trivial planes~\cite{szabo2020strain, Bernevig2020, Goswami2023}.
These higher-order topological planes support a quantized non-Abelian Berry flux~\cite{Goswami2023,tyner2024plane}, and may also be diagnosed by the filling  anomaly~\cite{Cano2021}.
In these materials, the band inversion parameter ($\Delta$) controls the separation between the DPs along the rotation axis, and tunes topological phase transitions characterized by a merging of the DPs. 
These considerations lead to the topological phase diagrams in Fig.~\ref{fig:DSMs}(a).

In cubic-symmetric DSMs the planes separating a pair of DPs along a fixed rotation axis can also be classified as higher-order topological insulators.
Interestingly, however, since cubic-symmetric DSMs possess multiple, mutually-orthogonal rotation-axes that host DPs, there exists \emph{interpenetrating} HOTI planes in the Brillouin zone, as depicted in Fig.~\ref{fig:bulk-topo}(a). 
Therefore, unlike rotation-symmetry protected DSMs, cubic-symmetry protected DSMs can not be considered as a stack of HOTI and ordinary insulator planes.
Instead, the bulk topology is inherently three dimensional, owing to its origin in the cubic point group. 

In order to explicitly demonstrate it, let us express the Hamiltonians introduced above in a schematic form,  
\begin{align}
H_0(\bs k) = N_0(\bs k) \mathbbm{1} + \vec N(\bs k) \cdot \vec \Gamma
\end{align}
where $\vec \Gamma$ collects the five mutually anti-commutating $\Gamma$-matrices.
%we compute the quantized non-Abelian flux expressed in terms of the relative or spin Chern number~\cite{Goswami2023}.
%%
On the $k_x-k_y$ planes, the  projections of the non-Abelian Berry curvature on the generators of $\hat C_{4z}$ rotations, $\Gamma_{12}$ and $\Gamma_{34}$, are given by~\cite{Goswami2023} 
\begin{align}
F_{xy}^{ab} = \sin\theta_{ab}\qty[\partial_{k_x}\theta_{ab} \partial_{k_y}\phi_{ab} - \partial_{k_y}\theta_{ab} \partial_{k_x}\phi_{ab}  ]
\end{align}
with $\theta_{ab} = \cos^{-1}\qty[1 - (N_a^2 + N_b^2)/|\vec N|(|\vec N| + N_5)]$ and $\phi_{ab} = \tan^{-1} N_b/N_a$, where we have suppressed the momentum dependence of $N_j$'s for notational convenience. 
%%%
These projected Berry curvatures support quantized Berry flux through the $k_z$ planes of the Brillouin zone, which is called the relative Chern number and it is equivalent to the spin Chern number~\cite{Goswami2023}.
While the net relative Chern number supported by $F_{xy}^{34}$ is identically zero, $F_{xy}^{12}$ supports
a quantized, $k_z$-dependent,  relative Chern number, 
\begin{align}
\mathcal C(k_z) = \frac{1}{2\pi} \int_{BZ} \dd{k_x} \dd{k_y} F_{xy}^{12}(\bs k),
\end{align}
as shown in Fig.~\ref{fig:bulk-topo}(b).
The corresponding results for the $sf$-DSM is presented in Appendix~\ref{app:sf-dsm}.
Since the number of HOTI planes separating the pair of DPs on a fixed rotation axis is controlled by $\Delta$, the average value of $\mathcal C$, 
\begin{align}
\mathcal C_{\text{avg}}(\Delta) = \int_{-\pi}^\pi \frac{\dd{k_z}}{2\pi} \mathcal C(k_z, \Delta),
\end{align}
allows us to locate the topological phase boundaries along $\Delta$, as shown in the inset of Fig.~\ref{fig:bulk-topo}(b).
An unquantized but finite value of $\mathcal C_{\text{avg}}(\Delta)$, with $\mathcal C(k_z, \Delta)$ being quantized and non-zero for a subset of $k_z$, is the signature of a DSM phase. 
By contrast, the ordinary and topological insulators obtained within our models exhibit a quantized value of $\mathcal C_{\text{avg}}(\Delta)$, with $|\mathcal C_{\text{avg}}(\Delta)| = 0$ ($1$) in an ordinary (topological) insulating phase.
This behavior of $\mathcal C_{\text{avg}}(\Delta)$ is reflected by the topological phase diagrams in Fig.~\ref{fig:DSMs}(b). 
We note that the phase diagrams are symmetric about $\Delta = 0$ because the term multiplying $\Gamma_5$ in $H_0^d$ and $H_0^f$ changes its overall sign under $\Delta \mapsto -\Delta$ and $k_j \mapsto k_j + \pi$. 
Although this mapping changes the location of the axes on which the Dirac points lie, the topology of the phases related by a change of sign of $\Delta$ does not change qualitatively. 
Henceforth, without loss of generality, we will focus on $\Delta \geq 0$.

As depicted in Fig.~\ref{fig:DSMs}(b), the four-band model of $sd$-DSMs permits two distinct DSM phases: ``DSM-1'' for $1<\Delta <3$ and ``DSM-2'' for $\Delta <1$. 
The DPs in the DSM-1 (DSM-2) phase lie along the principal axes on the $k_j = 0$ ($k_j = \pi$) planes, which results in a total $2 \times 3 = 6$ ($4 \times 3 = 12$) DPs.
While both DSM phases are examples of cubic-symmetric DSMs, here, we focus on the DSM-1 phase for its relevance to recent experiments in  REPns which we will discuss below.

Since a cubic crystal possesses multiple threefold and fourfold rotation axes, relative Chern numbers can be defined by projecting the non-Abelian Berry curvature on the respective rotation generators, which will topologically classify the planes perpendicular to the  rotation axes.
Here, $F_{yz}^{13}$ (projected on the $\hat C_{4x}$ generator $\Gamma_{13}$) and $F_{zx}^{23}$ (projected on the  $\hat C_{4y}$ generator $\Gamma_{23}$) support non-trivial relative Chern numbers in analogy to $F_{xy}^{12}$, and their jumps along the $k_x$ and $k_y$ axes, respectively, can be used to identify the Dirac points lying along these axes as Berry curvature monopoles. 
Since the three rotation generators supporting non-trivial relative Chern numbers do not commute, these monopoles are associated with different sectors of the non-Abelian Berry gauge group, which are related by the cubic symmetry.
Thus, a specific projected Berry flux will not detect all three pairs of DPs.
We note that this is a unique feature of cubic-symmetric DSMs and it is absent in DSMs arising in trigonal/tetragonal/hexagonal crystals.
We summarize this feature through Fig.~\ref{fig:bulk-topo}(c), where we plot the local texture of the non-Abelian Berry curvature projected on the aforementioned rotation generators.

\begin{figure*}[!]
\centering
\subfloat[]{%
\includegraphics[width=0.37\linewidth]{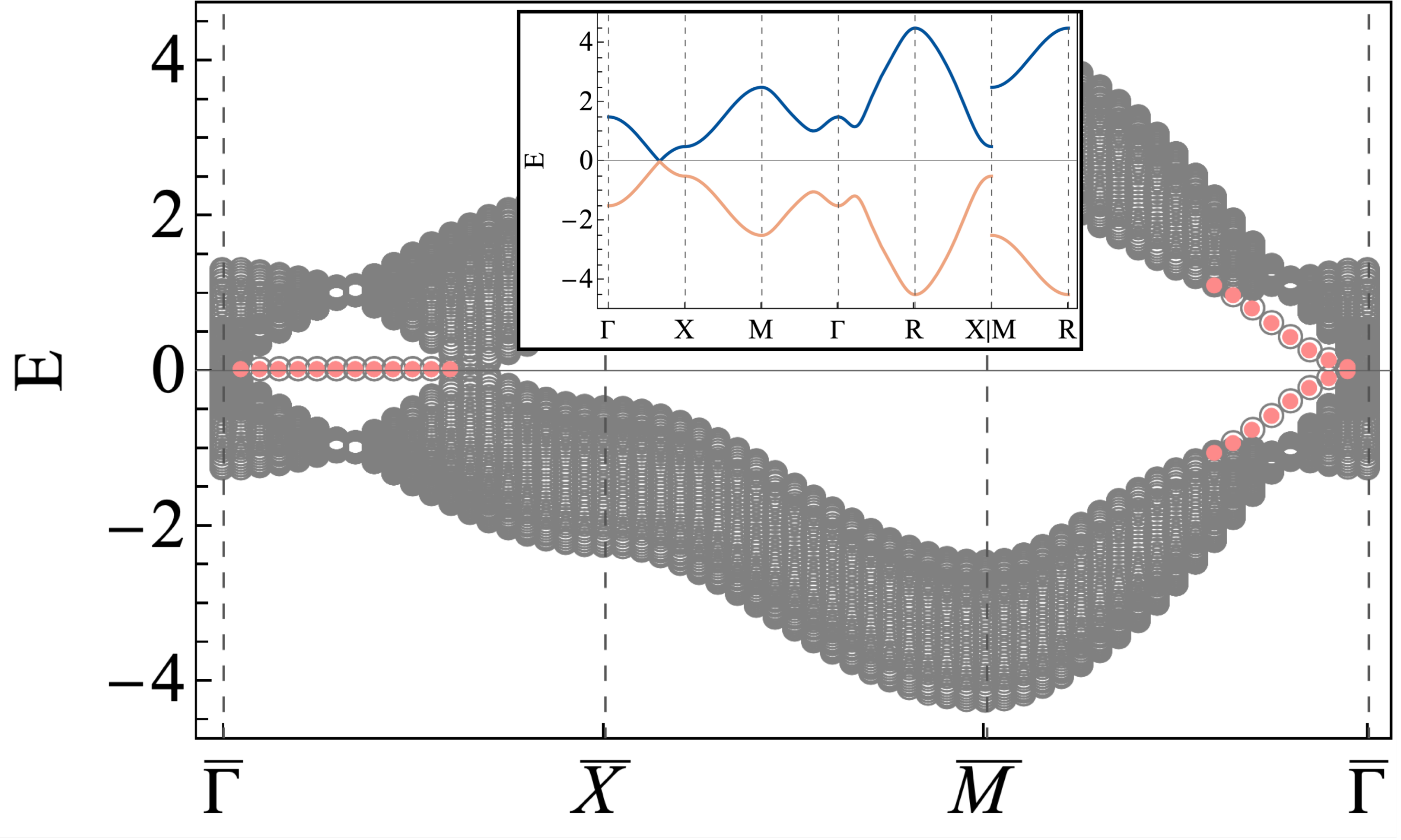}%
}
\hfill
\subfloat[]{%
\includegraphics[width=0.27\linewidth]{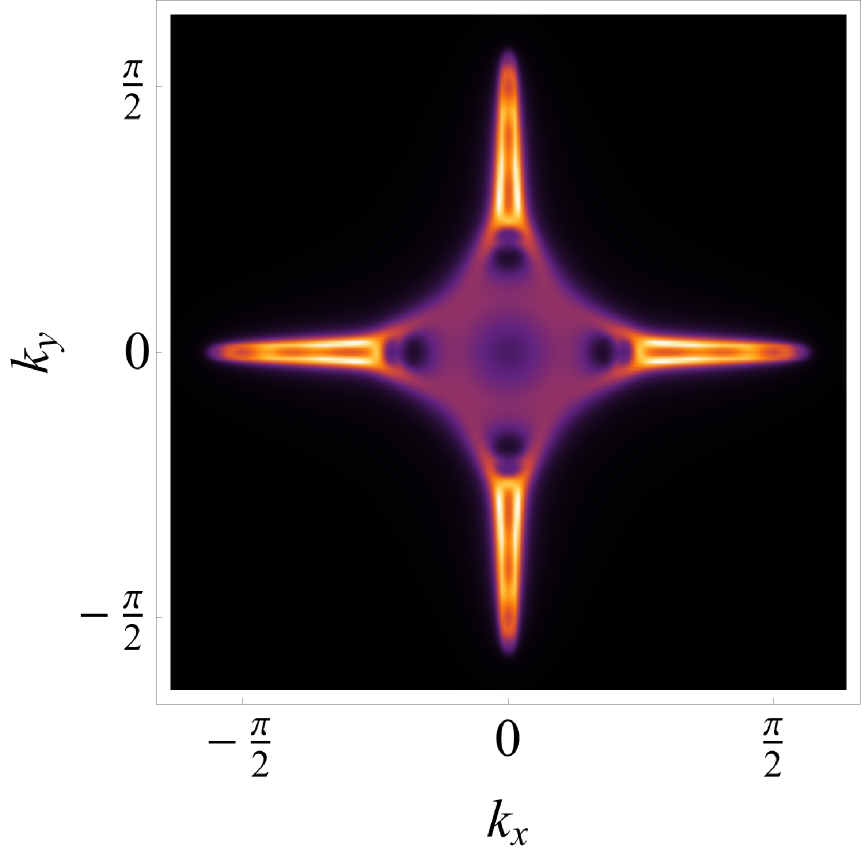}%
}
\hfill
\subfloat[]{%
\includegraphics[width=0.33\linewidth]{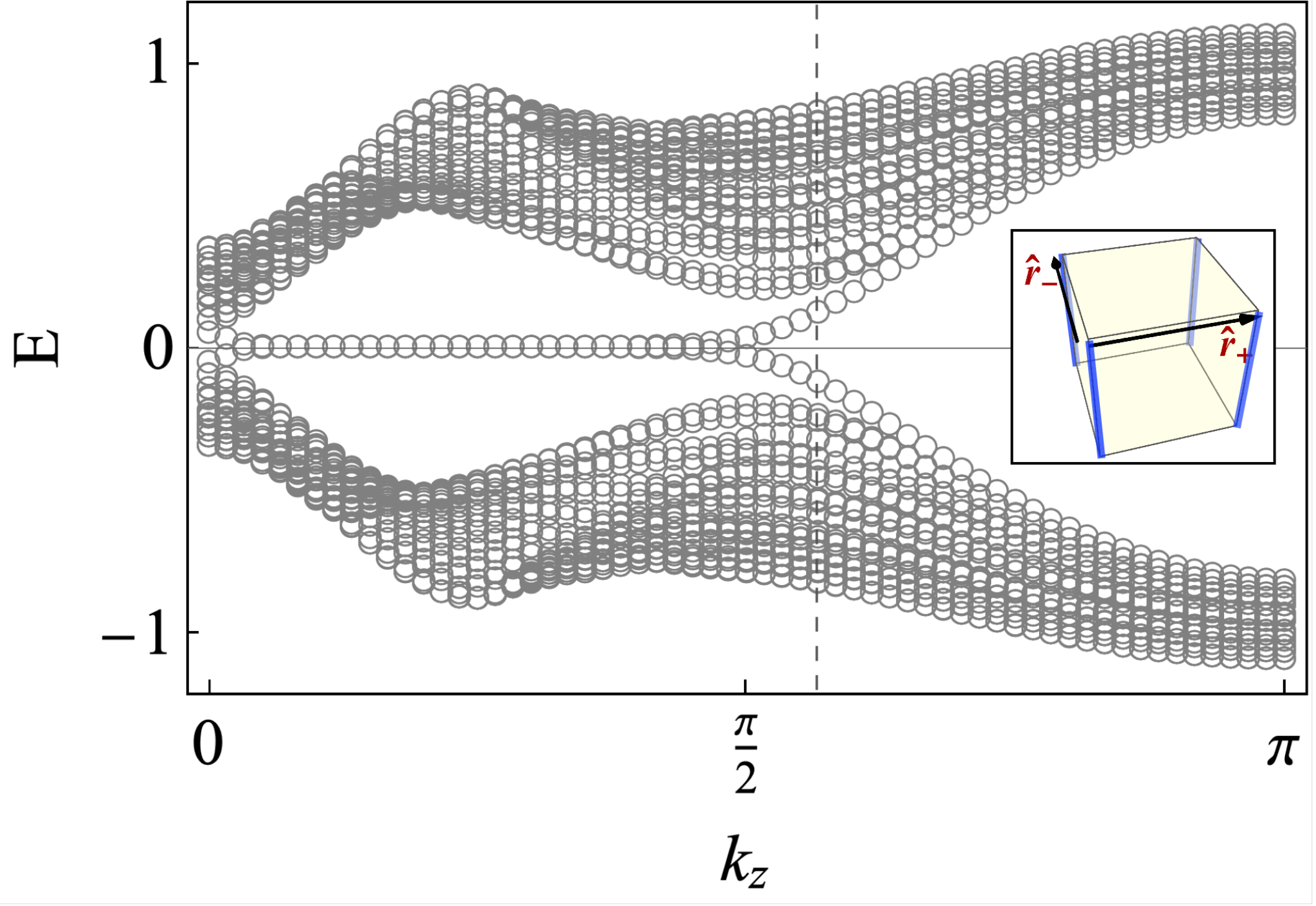}%
}
\hfill
\subfloat[]{%
\includegraphics[width=0.37\linewidth]{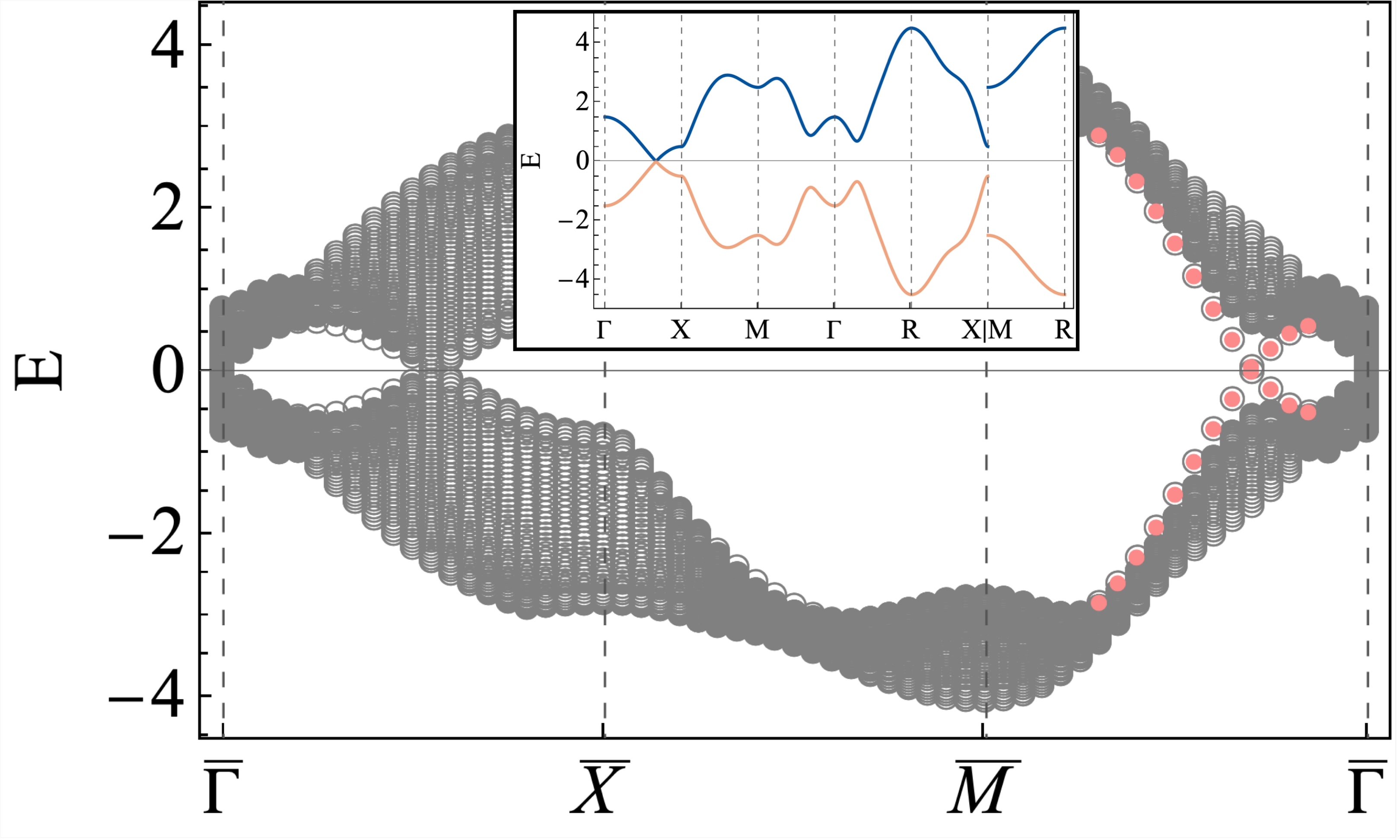}%
}
\hfill
\subfloat[]{%
\includegraphics[width=0.27\linewidth]{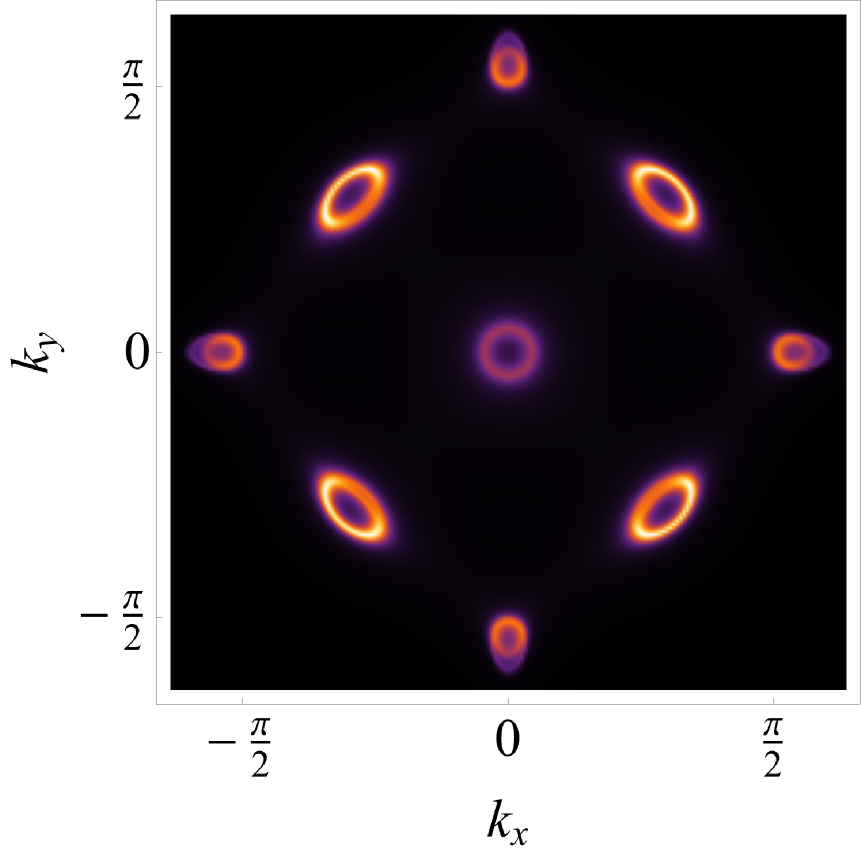}%
}
\hfill
\subfloat[]{%
\includegraphics[width=0.33\linewidth]{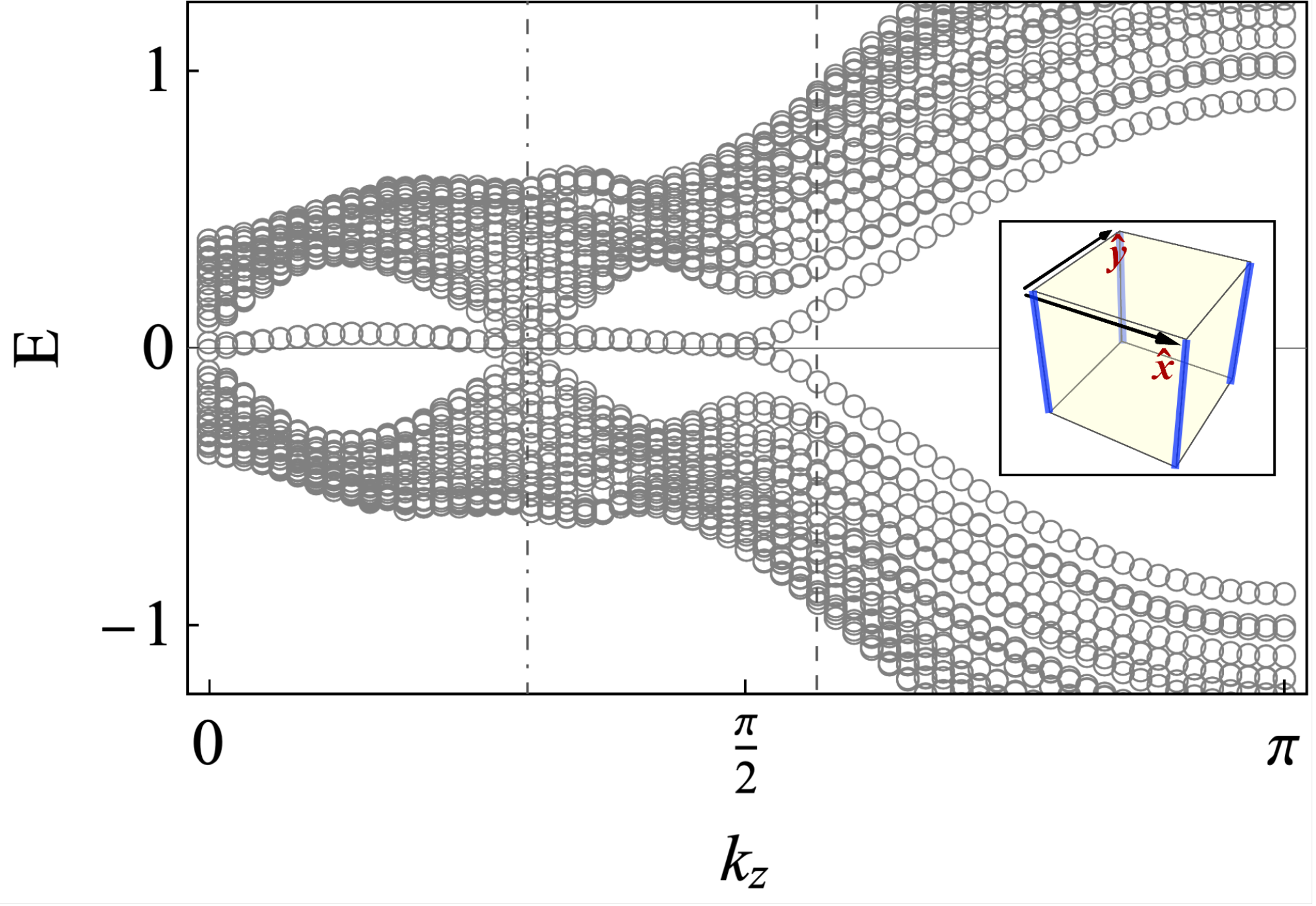}%
}
\hfill
\caption{Surface and hinge states of the s-d and s-f DSM. 
Gapless states on the (001) surfaces of (a) s-d and (d) s-f DSM (red discs).
In the inset, the bulk bandstructure is presented for comparison, which shows the bulk DP along the $\Gamma-X$ line.
Fermi surfaces supported by the surface states at low surface-dopings in  
(b) s-d DSM at $\mu/t_0 = 0.05$ (e) s-f DSM at $\mu/t_0 = 0.15$.
The surface-Fermi surfaces (SFS) are incomplete, i.e. form arcs, for s-d DSMs. 
By contrast, in s-f DSMs the SFS's may form closed Fermi pockets.
Hinge states on hinges along $\hat z$ (highlighted in blue) and perpendicular to  (c) the [100] direction in s-d DSM and (f) the [110] direction in s-f DSM.
The vertical dashed  lines in (c) and (f) [dot-dashed line in (f)] mark the location of projection of the bulk DP [surface DP].
Here, we have used $(t_1^{d/f}, t_3^{f},  \Delta, w_1^{d/f}, w_2^{d/f}) = (1.2, 1.2, 1.8,0, 0)$ and $t_2^{d/f} = 1$, and defined $\hat r_\pm = (\hat x \pm \hat y)/\sqrt{2}$. 
}
\label{Fig2}
\end{figure*}

%%%%%%%%%%%%%%%%%%%%%%%%%%%%%%%%%%%%
\section{Boundary-localized states}
We now present our results on the boundary localized states in two subsections. 
First, we  discuss properties of the non-magnetic parent phase, followed by the introduction of magnetism and its potential relevance to rare-earth magnets.

\subsection{Non-magnetic phase} 

Here, we study the various bulk, surface and hinge state properties of the two model Hamiltonians Eq.~\ref{Eq:HdHf}. Fig.~\ref{Fig2} panels (a-c) and (d-f) show the bulk and surface band structures, surface Fermi surface, and hinge states of $H_0^d(\mathbf{k}), H_0^f(\mathbf{k})$ respectively.  Since the off-diagonal terms in both Hamiltonians vanish along all $\Gamma$-$X$ high-symmetry lines in the Brillouin zone, the DPs are located at the intersections of these lines and a 2D surface defined by the condition:
\[
\sum_{j=x,y,z} c_{k_j} = \Delta.
\]
For $1 < \Delta < 3$, this surface is a spheroid centered at the $\Gamma$ point. In this parameter regime, six DPs emerge, two along each principal axis. A plot of the bulk band structure for both models is shown in the inset of Fig.~\ref{Fig2} panels (a) and (d).  These DPs are protected by four-fold rotations about the three co-ordinate axis. In both models, the pair of degenerate bands along the $\Delta$ line transform as $\bar{\Delta}_6(2)$, $\bar{\Delta}_7(2)$ (see Fig.~\ref{Fig1}). Hence, the crossings are protected by the distinct rotational eigenvalues $e^{\pm i \frac{3 \pi}{4}}$ and $e^{\pm i \frac{\pi}{4}}$ of four-fold rotations. 
%%%%%%%%%%%%%%%%%%%%%%%%%%%%%%%%%%%%

\begin{figure*}[!]
    \centering
    \includegraphics[width=\textwidth]{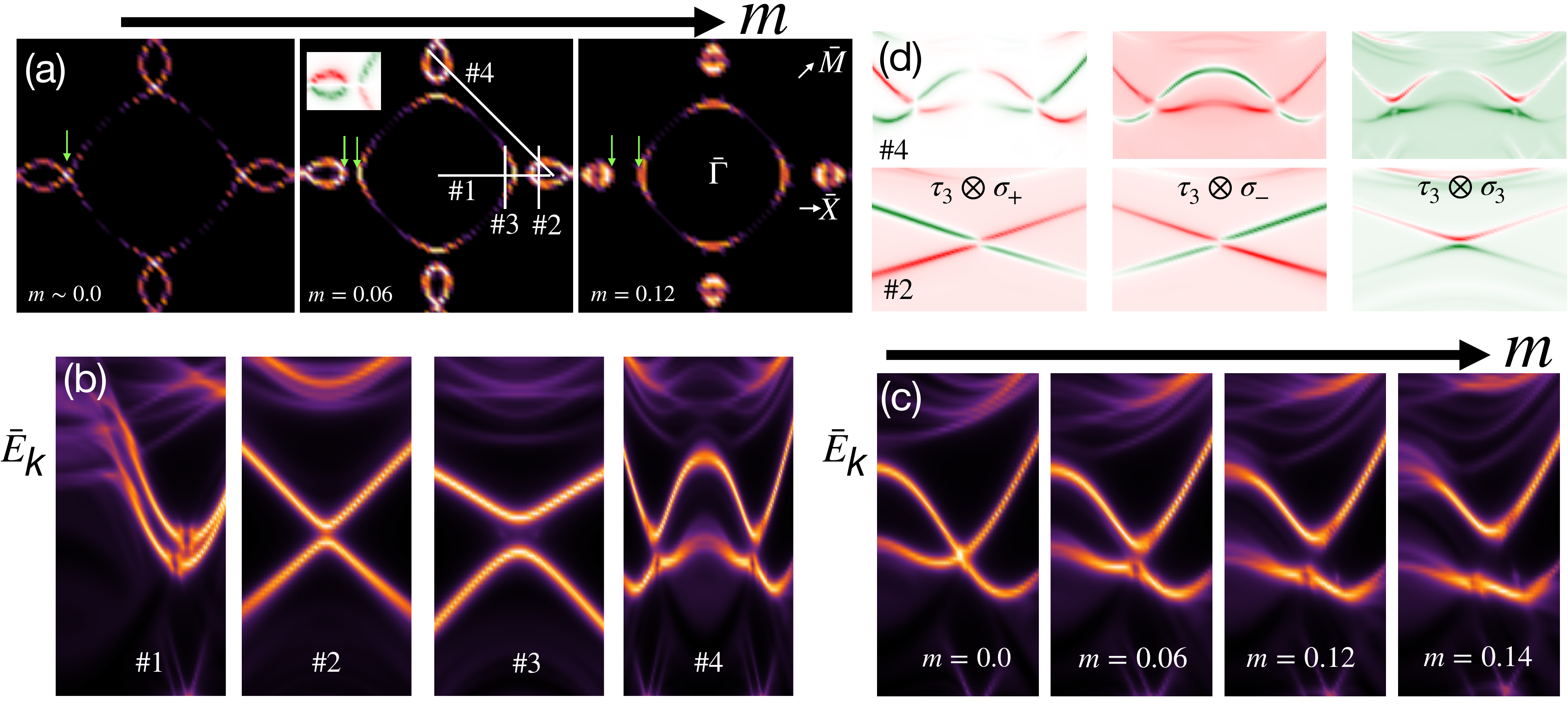}
    \caption{ The (001) surface state structure of $s$-$d$ DSM in the presence of broken time-reversal symmetry. Panel (a) shows the spectral function of the  ``surface" Fermi surface as a function of the magnetic order parameter $m=m_1=m_2$. The green arrows mark splitting of the electron pocket and hole-like Fermi arc. Inset within the $m=0.06$ panel shows spin texture for in plane spin polarization. Panel (b) shows $\omega-\bs k$ spectral function plots along different cuts $\#1-\#4$ shown in panel (a). Panel (c) shows the evolution of the spectral function along cut $\#4$ with increasing strength of magnetization in the magnetically ordered phase. 
    The polarization-dependent spectral function for cuts $\#4$ and $\#2$ appears in panel (d) for in-plane ($\sigma_{\pm}$) and out-of-plane ($\sigma_3$) polarizations marked in individual panels. The parameters chosen are $t_1^d=0.9, t_2^d=1, w_1^d=0, w_2^d=-1.4, \Delta=1.1, \mu=0.15 $. The value of $m$ is set to $0.07$ for panels (b), (d) and is marked in the various sub-panels for (a) and (c). }
    \label{RE-Pn-SurfaceStates}
\end{figure*}
%%%%%%%%%%%%%%%%%%%%%%%%%%%%%%%%%%%%

Next, we show the surface state band structure in the main plots of Fig.~\ref{Fig2} (a,d) for the two models on the equivalent (001) surfaces. For the $s$-$d$ case, the surface state dispersion can be understood as a deviation from perfectly flat drumhead states, which would arise if the $\sin{k_x} \sin{k_y}$ term in the lattice Hamiltonian were set to zero. 
In that case, a nodal loop forms in the $k_z = 0$ plane (assuming $1 < \Delta < 3$), which supports the drumhead surface state on the (001) surface. 
However, introducing the finite $\sin{k_x} \sin{k_y}$ term adds dispersion to the drumhead, leading to the surface Fermi surface states shown in Fig.~\ref{Fig2}(b). 
Thus, the dispersion is given by $E_\pm(k_x, k_y) \sim \pm A |\sin{k_x} \sin{k_y}|$, where $A$ is  a model-parameter dependent constant. 
The surviving gaplessness of these states along $\bar \Gamma - \bar X$ is a consequence of the space-group symmetries which  constrain the form of the term that gaps the drumhead states.
Since a subset of the $k_z$-planes in $sd$-DSMs are 2D HOTIs, we expect to find hinge localized states.
In fact, these 2D HOTI planes are  related by a basis transformation to  2D second order topological insulators~\cite{benalcazar2017quantized}, and support corner-localized states in  a diamond geometry~\cite{sur2022mixed}.

Therefore, we focus on the  hinges that occur at the intersection of [110] and [1$\bar{1}$0] surfaces parallel to the $\hat z$ direction, and obtain the spectrum of the hinge-localized states, as shown in  Fig.~\ref{Fig2}(c).
These hinge states are degenerate due to the four-fold rotational symmetry about the z-axis, and they are guaranteed to be mid-gap states by the $\mathbb Z_2$ chiral symmetry of $H_0^d$~\cite{benalcazar2017quantized}. They further connect the projection of the bulk Dirac crossings along the $\hat k_z$ direction. 
We note that, here, the chiral symmetry is physical, as detailed in section~\ref{sec:models}, and not \emph{ad hoc}.

For the $s$-$f$ case, the absence of a space-group-enforced chiral symmetry generally disallows degeneracy of states on the equivalent [001] surfaces [Fig.~\ref{Fig2} (d); see also Appendix A]. 
As a result, a pair of non-degenerate, spin polarized surface states emerge from the projection of DPs on the surface $\bar{\Gamma}$ point and merge into the projection of the DPs along the $\Gamma$-$X$ line. 
However, along the $\Gamma$-$M$ line, these surface states cross with the crossing protected by the mirror symmetry perpendicular to the [110] axis. 
These crossings are topologically protected since the $k_x = \pm k_y$ planes in the Brillouin zone support a mirror-Chern number of magnitude $2$.  
The surface Fermi surface correspondingly contains small electron or hole pockets along the $\Gamma$-$M$ direction, as shown in Fig.~\ref{Fig2} (e) depending on the filling constraint. 
The parity and orbital geometry of the $s$-$f$ DSM allows hinge states along the $\hat z$ hinge perpendicular to the [110] direction  rather than the [100] direction. 
The hinge states connect  projections of both the bulk and surface DPs along the $k_z$ direction and they are generally dispersive [see Fig.~\ref{Fig2}(f)]. %as demonstrated in Fig.~\ref{Fig2}(f). 
\subsection{Magnetic phase} 
We now turn to the role of broken time reversal symmetry in the models in Eq.~\ref{Eq:HdHf}. As mentioned earlier, the exact nature of the magnetic space group in rare-earth magnets is uncertain. However, a relatively strong quenching of the magnetization below the magnetic transition temperature indicates a substantial anti-ferromagnetic component.  
We model the magnetization terms generically by adding the terms $\Gamma_{34}$, $\Gamma_{12}$ to $H_0^{d,f}(\bs k) $ %the non-interacting Hamiltonian,
\begin{align}
H(\bs k) &= H_0^{d,f}(\bs k) + H_M \\
H_M &= (m_1\Gamma_{34} + m_2 \Gamma_{12}),
\end{align}
where $m_1, m_2$ control the strength of the magnetic moments. These terms correspond to opposite and equal spin fields on the two orbitals/sublattices respectively. 
 We note that the chosen magnetic order does not break the translation symmetry of the parent state since the opposite spin field occurs between two sublattices within the same unit cell.

Previous studies on Weyl semimetals, including magnetic Weyl semimetals~\cite{Ong2016, Felser2018, YLChen2019, CK2022-Nat}, inversion-broken Weyl semimetals~\cite{Hasan2015, Hasan2015-2, YLChen2015, Ding2015, Kaminski2016, YLChen2017}, and systems with both magnetic and inversion-breaking orders~\cite{Hasan2018, White2020, Zhang2021, tyner2024dipolar}, have largely concentrated on tetragonal and hexagonal crystal structures. In contrast, Weyl semimetals derived from cubic symmetries are less explored but may emerge through a magnetic instability in various systems, such as ordinary insulators, topological insulators, Luttinger semimetals, or Dirac semimetals. In this study, we focus on the Dirac semimetal route to Weyl semimetals, which, as we will show, produces surface states that align well with recent ARPES observations. Notably, this approach offers a previously less examined pathway to Weyl semimetals, distinct from the more traditional tetragonal or hexagonal routes. Exploring this magnetic instability in cubic DSMs is crucial for understanding surface-state properties in rare-earth magnets with parent cubic symmetry, as it reveals how magnetism influences cubic Dirac semimetals and provides insights into the unique behaviors that may arise in these cubic systems. For the purpose of subsequent analysis,  we assume that DPs appear in proximity to the Fermi energy close to the magnetic transition. \textcolor{black}{However, for now,} we remain agnostic about how DPs emerge near the Fermi energy at the magnetic transition. Later in the paper, we discuss possible scenarios for how this could occur, whether through band renormalization due to correlations or spontaneous band reorganization caused by magnetism.

We now introduce the magnetic order terms $H_M$ to the cubic DSM. 
Upon breaking time-reversal symmetry, Dirac points generically split into Weyl points.
The resultant Weyl semimetal retains the higher order topological features of DSMs, and are examples of  higher-order Weyl semimetals~\cite{ghorashi2020higher, wang2020higher}. 
Here, we will encounter a higher-order Weyl semimetal in a cubic crystal, which is analogous to the Weyl semimetal obtained in the all-in-all-out state of pyrochlore irridates~\cite{witczak2012topological}.

Due to its potential relevance for rare-earth magnets, we focus on the $s$-$d$ Dirac Hamiltonian $H_0^{d}(\bs k)$ in this section and leave the discussion of $H_0^{f}(\bs k)$ to the Appendix.  \textcolor{black}{Henceforth, we set $t_1^{d/f}$ to be of order unity to obtain a reasonable bandwidth scale of $\sim 4-8$ times $t_1^{d/f}$.  The chemical potential and $w_1^{d/f}, w_2^{d/f}$  are chosen to obtain a surface fermi surface of size similar to that seen in ARPES experiments. }
\textcolor{black}{The effects of these magnetic terms  
  leading to the splitting of surface states are shown in Fig.~\ref{RE-Pn-SurfaceStates}(a-d) respectively (splitting of DPs into Weyl points in the bulk is shown in the Appendix [Fig.\ref{fig:sf-Chern}]).} We start with the total spectral function on the [001] surface. Due to the relatively small magnetic order parameter we choose, the side surfaces, [100] and [010], only weakly break C$_4$ symmetry as expected by the chosen magnetic order. As a result, they exhibit qualitatively similar features as the [001] surface. 
  The evolution of the surface Fermi surface for different values of the magnetization order parameter is presented in Fig.~\ref{RE-Pn-SurfaceStates}(a). \textcolor{black}{ We choose the parameters $m_1$ and $m_2$ to be small compared to the bandwidth.} 
  For close to zero magnetization, a weak intensity electron pocket touches an arc-like crossing at a relatively higher-intensity vertex marked by a green arrow. 
  Upon increasing the magnetic order parameter (with $m=m_1=m_2=0.06$), a high intensity surface Fermi surface is formed where the electron pocket splits off from a hole-like Fermi arc (green arrows). The hole-like Fermi arc is centered around the surface $\bar{\Gamma}$ and electron-like pockets are located along the $\bar{\Gamma}-\bar X$ line toward the $\bar{X}$ point. The Fermi arcs merge into the bulk along the $\bar{\Gamma}-\bar M$ line; hence, there is no spectral intensity along the surface diagonals. Both of these features are consistent with ARPES observations~\cite{CK2022-Nat, CK2022-PRB}. Upon further increasing the magnetic order parameter ($m=0.12$), the splitting between the hole-like arc and electron pocket further increases (green arrows) also observed in ARPES measurements~\cite{CK2022-Nat, CK2022-PRB}. With an increasing chemical potential (not shown), the electron pockets increase in size whereas the hole-like features shrink while maintaining the four-fold symmetric Fermi arcs. 
  As a reference, several momentum space cuts in the $k_x-k_y$ plane of the surface Brillouin zone are labeled $\#1-\#4$ in Fig.~\ref{RE-Pn-SurfaceStates} (a). The surface band structure along these cuts is illustrated in Fig.~\ref{RE-Pn-SurfaceStates}(b) and described below. Here, $\bar E_{k}$ denotes the surface energy dispersion. \textcolor{black}{Finally we note that for very large magnetization any non-trivial semimetallic topology is lost as the band crossings are absent. }

Features that develop into hole-like arcs and electron pockets 
are driven by magnetic order. In the absence of magnetic order, the surface bands are fully degenerate along cut \#1. Introduction of magnetism leads to \textit{spin-splitting} of these degenerate bands. For larger magnetization, the spin-split bands eventually form the hole-like Fermi arcs and electron-like pockets [this must be compared with the $s$-$f$ DSM (see Appendix) where splitting occurs already in the non-magnetic phase and does not evolve analogously along cut \#1]. On the other hand, along cuts \#2--\#4, the generic Dirac Hamiltonian $H_0^d(\bs k)$ produces spin-split bands with opposite curvature even without magnetism. These bands, however, touch at a single point along each of these cuts. When a non-zero magnetic moment is introduced, a \textit{gap opening} occurs at these band-touching points. The features along cuts \#1 and \#2--\#4, after the spin-splitting and gap opening respectively, are shown in Fig.~\ref{RE-Pn-SurfaceStates}(b) and align with the ARPES data reported in ~\cite{CK2022-Nat, CK2022-PRB}. As an example of how the gap opening evolves as a function of magnetization strength, we plot the surface spectral function along cut \#4 as a function of the magnetization strength in Fig.~\ref{RE-Pn-SurfaceStates}(c). When the magnetization is zero, the bands are degenerate at a single point along cut \#4. Upon introducing magnetism, the degeneracy opens up a gap into hole-like and electron-like bands. The gap subsequently increases as the magnetization increases consistent with ARPES observations~\cite{CK2022-Nat}.\par  We further study the spin texture of the surface state bands. Fig.~\ref{RE-Pn-SurfaceStates}(d) displays the spin-polarization-resolved surface spectral function along two of the cuts \#2 and \#4 indicated in Fig.~\ref{RE-Pn-SurfaceStates}(a). For the in-plane spin projections ($\sigma_{\pm}$), the gap opening leads to the formation of two hybridized surface state bands. For each band, the momentum where the gap opens  separates momentum regions where the bands exhibit opposite spin projections. This behavior occurs in both the electron-like and hole-like bands, though these two bands themselves have opposite overall spin projections. In this sense, there exists a non-trivial spin-momentum locking of the kind seen in the surface states of a topological insulator.  A similar bifurcation of the spin texture can be seen in the electron pocket and hole-arc in the surface Fermi surface (inset of Fig.\ref{RE-Pn-SurfaceStates}(a), $m=0.06$). The top and bottom sections of the  electron pocket and hole-arc have opposite in-plane spin polarization.
In contrast, for out-of plane spin projection ($\sigma_3$ panel of Fig.\ref{RE-Pn-SurfaceStates}(d)), while the gap opening still produces two bands with opposite spin polarization, each band maintains mostly uniform spin polarization throughout for all momenta. These polarization dependent calculations are consistent with observations made by spin-resolved ARPES measurements in the rare-earth monopnictide magnets~\cite{CK2022-Nat}.\par

\section{Discussion and Outlook}
The results presented above suggest that several key experimental characteristics observed in the rare-earth magnets may be attributed to $s$-$d$ type DSM physics. Observations such as surface state dispersions along high-symmetry axes, surface Fermi surface topology--featuring hole-like Fermi arcs at the \(\bar{\Gamma}\) point and electron pockets near \(\bar X\) along the \(\bar{\Gamma}-\bar X\) direction -- gap openings, and polarization dependence all suggest DSM physics as a potential origin of the observed phenomena. 
However, it remains unclear whether the Dirac crossing responsible for these properties originates from the parent compound or emerges due to the magnetic space group. 
DFT calculations of the parent bulk electronic structure show both topological insulator (TI) band repulsion and Dirac crossings along the C$_4$ rotation axes~\cite{CK2023-CP}, but these DPs are located far from the Fermi energy (around 1 eV). \textcolor{black}{ In the presence of moderate to strong electron correlations, the band structure can undergo non-rigid modifications, primarily manifesting as band renormalization and/or quasiparticle damping, which reduces the sharpness of the bands. Band renormalization, characterized by the quasiparticle weight factor \( Z_k \), can lead to a reduction in the overall bandwidth, potentially pushing Dirac cones closer to the Fermi energy~\cite{liao2023orbital, chen2024emergent, Si-Lai-PNAS-2018, Si-NatPhys-2022, Si-Hu-2021, Si-Setty-2024}. However, we note that not all members of the RE-Pn family exhibit strong correlations; some are only weakly correlated and, consequently, show minimal band renormalization effects.} Additionally, there is little experimental evidence of surface states along certain directions \textit{above} the Neel transition. If the parent Dirac crossing is responsible for these features, magnetic order must significantly renormalize the parent state bands. These unresolved issues raise important questions about the role of the parent DSM in shaping the surface state properties, which warrant further investigation. \par

Alternatively, the Dirac crossings could arise \textit{spontaneously} from the low-temperature magnetic space group, though the specifics of this space group for these materials remain unresolved. This scenario explains the absence of surface state features above the Neel transition. 
Additionally, the possibility of a transition from a cubic parent phase to a fully tetragonal magnetic space group could account for the large anisotropy between the top and side surfaces seen in experiments.
In systems with single wave-vector AFM orders, DPs are aligned with a single C$_4$ rotation axis, which would result in no surface states on the plane perpendicular to this axis, while the side surfaces would still support surface states.
Due to the small out of plane magnetic order used in our calculations, the top and side surfaces are only weakly anisotropic and cannot explain the presence of surface states only on certain surfaces. On the other hand, this scenario requires significant band folding along the magnetic ordering direction, which has little experimental support~\cite{CK2022-Nat}. 
Moreover, while magnetic space groups can protect DSMs via non-symmorphic time-reversal symmetries, these symmetries also impose constraints on surface state splitting. 
For instance, if the magnetic space group allows for a large surface-state splitting along the $\bar{\Gamma}$-$\bar X$ axis (as in the $s$-$f$ DSM) immediately below the magnetic transition, it is hard to explain how the separation of hole and electron pockets (e.g., along cuts \#1 and \#4) is proportional to the magnetic order parameter. 
Similarly, if the magnetic space group enforces surface state degeneracy along the $\bar{\Gamma}$-$\bar X$ axis for all temperatures below the magnetic transition, it cannot account for the observed magnetic order parameter-driven band splitting along this direction.

Several additional key points are worth noting. First, the splitting between the hole Fermi arc and the electron pocket on the surface Fermi surface starts at zero and gradually increases, correlating with the strength of the magnetic moment. This suggests that the splitting along the $\bar{\Gamma}$-$\bar X$ direction also evolves gradually with the magnetic order, indicating that the effective DSM description is likely of the $s$-$d$ type rather than the $s$-$f$ type, although a higher-band $s$-$p$ model cannot be ruled out. Second, in cubic environments, magnetic evolution may cause spin-orbit coupling (SOC) effects that are subtle and difficult to resolve experimentally. Therefore, ARPES experiments along non-high-symmetry directions have to be interpreted with care as it is important to categorize the \textcolor{black}{and Kaminski-Canfield spectral gaps~\cite{CK2022-Nat, CK-2024CommMat, CK-2024PRB} into spin-splitting and gap-opening scenarios.} In the former, bands are initially degenerate across an entire momentum cut, while in the latter, degeneracy exists only at specific points. Our calculations along cut \#4 (Fig.~\ref{RE-Pn-SurfaceStates}) confirm this is a gap-opening scenario, where spin-orbit coupling results in a finite number of band-touching points \textit{even before} the transition into the magnetically ordered phase occurs. Although less relevant for the rare-earth monopnictide magnets, a spin-splitting scenario with bands having opposite curvature is in principle allowed by other unconventional orders  
with non-trivial transformation properties of magnetic order parameters as is the case in altermagnetic ground states.

Third, Fermi arc-like features can emerge when DPs split into Weyl points along or perpendicular to the rotation axis. 
In this case, the splitting occurs along the axis, leading to a hole-like pocket that becomes buried in the bulk states along the $\Gamma$-M direction, producing the arc-like features observed experimentally. Fourth, if DSM physics is indeed essential for understanding the complex topological properties of magnetic order in RE-Pns, as our analysis suggests, then the presence of hinge modes would serve as a key prediction. Observing these hinge modes through probes such as scanning tunneling microscopy would confirm the key role of DSMs as the origin of various topological properties in RE-Pns. Finally, our model relies on two effective sublattices with spins in a cubic environment with Dirac crossings protected by the intersection of $\bar{\Delta}_{6}(2)$ and $\bar{\Delta}_{7}(2)$ representations. These conditions can be satisfied by other material candidates and space groups as well.  
We propose rare-earth auricuprides as promising candidates, given their antiferromagnetic nature and simple cubic structure with DPs closer to the Fermi energy. 
Additional candidate materials for realizing cubic-symmetric DSMs are Sn$_2$Ir and Rh$_2$As.
Further investigation of these materials is warranted to confirm these predictions. \par

In conclusion, we have developed minimal models for DSMs protected by cubic lattice symmetries, focusing on systems with two sublattices and incorporating potential magnetic orders. This framework allows us to explore bulk, surface, and hinge properties, including surface state dispersion, magnetic Fermi arcs, polarization effects, band splitting, and hinge modes.
Despite the lack of a robust bulk-surface correspondence in DSMs, our models provide valuable insights into surface state phenomena observed in magnetic topological semimetals, such as rare-earth monopnictides, addressing several key questions and discussing unresolved issues.  The presence of hinge modes provides a key prediction of DSM driven phenomenology in these materials. More specifically,  even-orbital $sd$-DSMs show an effective $\mathbb{Z}_2$  chiral symmetry that has key consequences for surface and hinge-localized states. Furthermore, our models offer a framework for exploring correlation effects in cubic Dirac systems, and we have identified other classes of potential cubic parent materials for further experimental validation. These findings highlight that DSM physics not only sheds light on diverse material behaviors but also opens new opportunities for exploring topological properties in magnetic semimetals. Future work should focus on experimentally testing these predictions to further refine our understanding of DSM behavior in cubic systems. \\ \newline
$\dagger$ csetty@iastate.edu
\paragraph*{{\bf Acknowledgements:}}
We extend our gratitude to P. Canfield and A. Kaminski for bringing these systems to our attention and for their valuable insights regarding the ARPES results. We also acknowledge helpful discussions with R. Flint, R. Prozorov, L.L. Wang, A. Tyner, and Y. Fang. 
S.S. is supported by NSF (DMR-2220603) and AFOSR (FA9550-21-1-0356). C.S. acknowledges support from Iowa State University and Ames National Laboratory start up funds.

%%%%%%%%%%%%%%%%%%%%%%%
\clearpage

\appendix
\onecolumngrid
\section{} \label{app:sf-dsm}
\begin{figure}[!t]
\centering
\subfloat[]{
\includegraphics[width=0.45\linewidth]{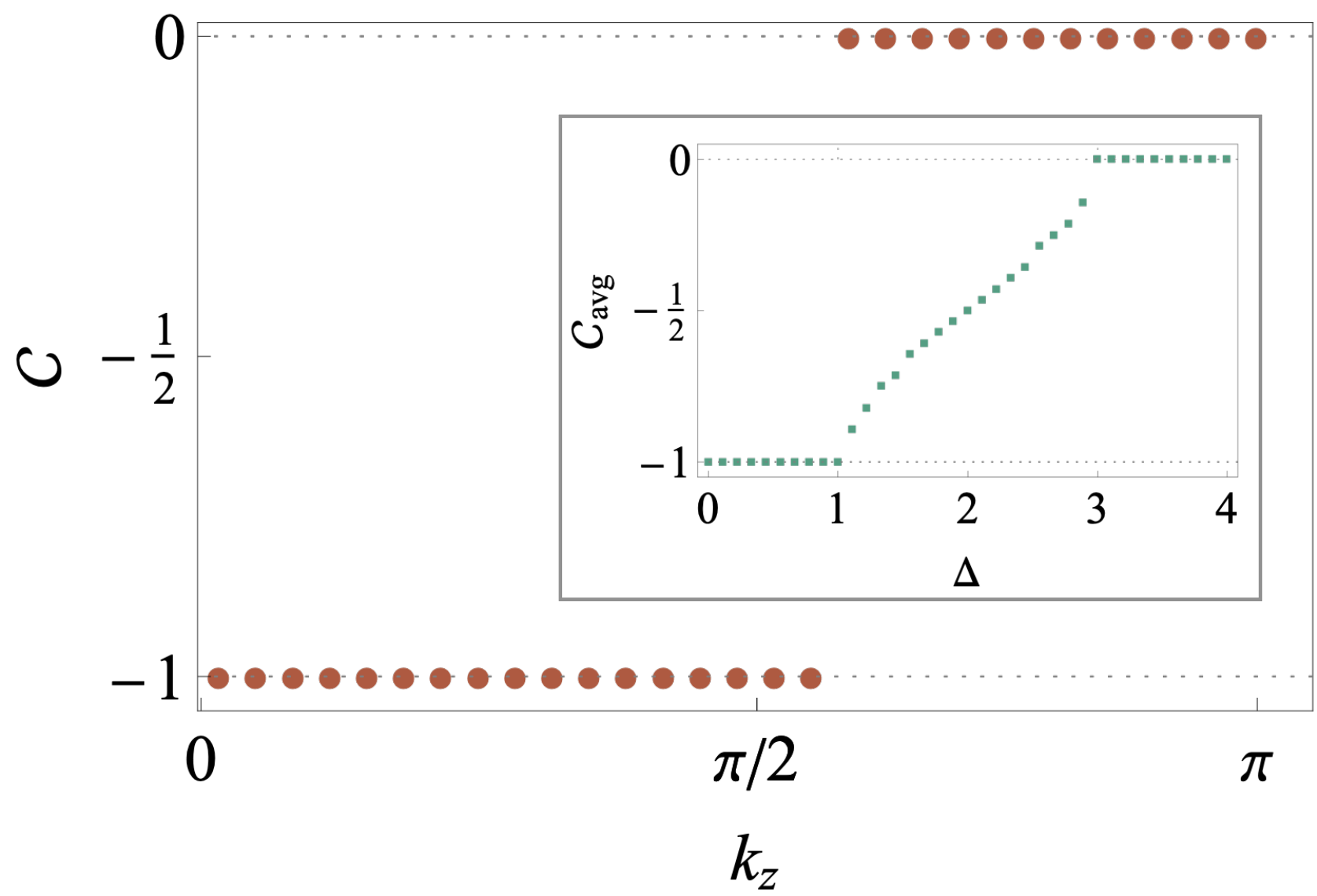}}
\hfill
\subfloat[]{\includegraphics[width=0.45\linewidth]{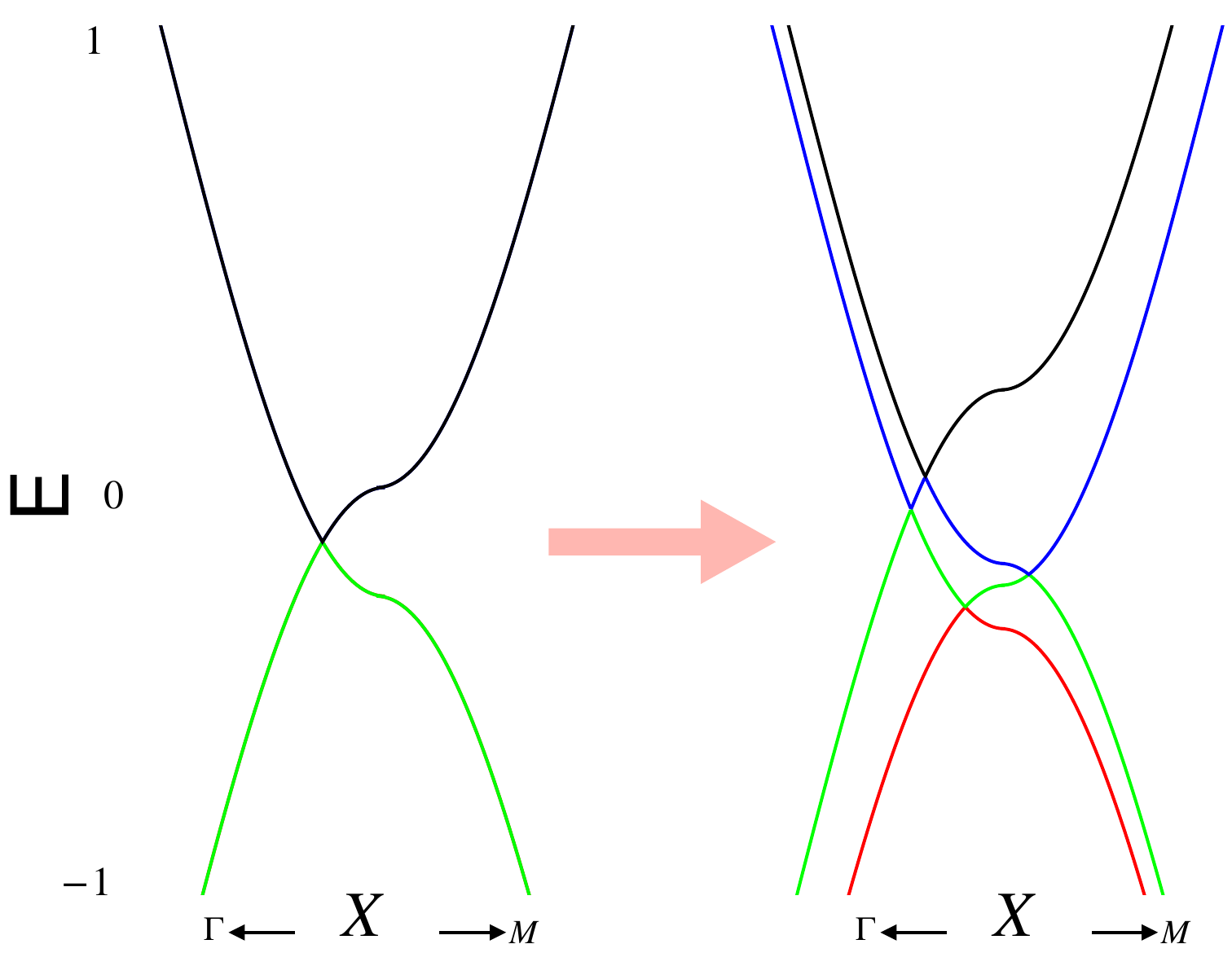}}
\caption{(a) Quantized flux (expressed in terms of the relative Chern number, $\mathcal C$) supported by the planes in $sf$-DSM. (inset) A characterization of the $\Delta$-driven phase diagram using the average $\mathcal C$, as discussed in the main text. 
\textcolor{black}{For both figures we have used $(t_1^f, t_2^f, t_3^f, w_j^d, \mu) = (1.4, 1.0, 1.2, 0.0, 0.0)$, while for the main figure we have also fixed $\Delta = 1.8$.}
\textcolor{black}{(b) Bulk splitting of the DSM into Weyl points with the introduction of magnetic order. The parameters chosen are $t_1^d=0.9, t_2^d=1, w_1^d=0, w_2^d=0, \Delta=1.1, \mu=0.15 $, $m_1=0.06$, $m_2 =0.1$. Note that we have set the tilt term to zero compared to the parameters used in the main text to better highlight the splitting effect. } }
\label{fig:sf-Chern}
\end{figure}

\textit{More on $sf$ Dirac semimetals:} In order to calculate the relative Chern number, we note that the generators of fourfold rotation  about $\hat z$, $\Gamma_{12}$ and $\Gamma_{34}$, are identical to that for rotation-symmetry protected Dirac semimetals discussed in Ref~\cite{Goswami2023}. 
Therefore, we follow the steps outlined therein to obtain the results in Fig.~\ref{fig:sf-Chern} (Left) and also the analogous figure for $sd$-DSM in the main text.
The quantized value of $\mathcal C_\text{avg}$ (see inset) in the region $0<\Delta<1$ indicates the presence of a topological insulator. 
By contrast, the interpolating behavior for $1<\Delta<3$ indicates the cubic-symmetric DSM phase studied in this paper. 

\begin{figure*}[!]
    \centering
    \includegraphics[width=\textwidth]{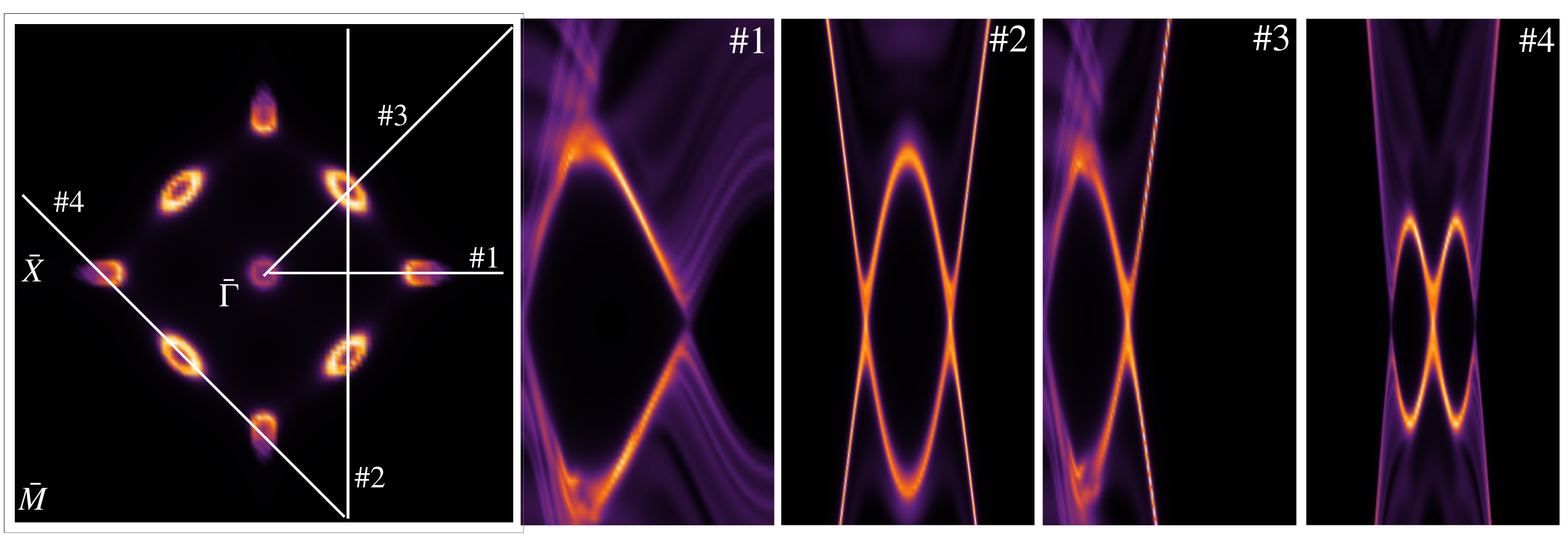}
    \caption{The (001) surface state spectral function of $s-f$ DSM in the presence of a small broken time reversal symmetric order parameter. The leftmost panel shows the surface Fermi surface, and the different energy cuts labeled $\#1-\#4$ are shown in panels labeled accordingly.  The parameters chosen are $m_1 = m_2 = 0.03, (t_1^{f}, t_3^{f},  \Delta, w_1^{f}, w_2^{f}, \mu) = (1, 1, 1.5, 0, 0,0.2)$ and $t_2^{f} = -1$.   }
    \label{SMFigSF}
\end{figure*}

The spectral function for the (001) surface under weak magnetic order is depicted in Fig.~\ref{SMFigSF} for the parameters chosen as $m_1 = m_2 = 0.03$, $(t_1^{f}, t_3^{f}, \Delta, w_1^{f}, w_2^{f}, \mu) = (1, 1, 1.5, 0, 0, 0.2)$, and $t_2^{f} = -1$. In the leftmost panel, the surface Fermi surface reveals electron pockets originating from Dirac cones positioned along the diagonal directions. Notably, an electron pocket with comparatively low intensity appears at the $\bar{\Gamma}$ point, where surface states merge with the bulk states, as shown in cut \#3.

Examining cut \#1, we observe that the surface states initially split into two branches, creating an increasingly large gap at small momenta. As momentum increases further, the gap gradually diminishes, and the surface states eventually merge into the bulk Dirac points. At intermediate momentum values, partially formed electron pockets emerge along the four equivalent directions of cut \#1. Importantly, this splitting of surface states is present right at the magnetic transition, indicating that the splitting along direction \#1 is very weakly dependent of the magnetic order parameter--unlike in the case of $s$-$d$ Dirac materials discussed in the main text where the surface state splitting is set by the magnetic order.

Cut \#2 shows two distinct electron pockets forming along the diagonal directions, where these pockets remain sharply defined as they cross without merging with adjacent bulk bands. Conversely, along cut \#4, the sharp surface states situated near the center of the cut gradually transition into the bulk states. In this process, they lose intensity and form electron pockets of relatively weak intensity along the $\bar{\Gamma}-\bar{X}$ direction. 

This configuration highlights the distinctive surface features of the spectral function in weak magnetic order, providing insight into the behavior of surface and bulk states across various directions in $s$-$f$ cubic Dirac semimetals.
\\ \newline
\textcolor{black}{\textit{DSM splitting:} 
 For completeness, we also examine the effect of the magnetization terms $H_M$ on the bulk band structure, as shown in Fig.~\ref{fig:sf-Chern} (Right). In general, the splitting of the doubly degenerate bands can occur in two distinct ways: either along the rotation axes or perpendicular to them. The terms proportional to $\Gamma_{34}$ and $\Gamma_{12}$ specifically induce splitting along the rotation axes. Currently, no bulk ARPES measurements have been reported, which will be crucial in clarifying the nature of the magnetic phase.}
%For completeness, we also show the effect of the magnetization terms $H_M$ on the bulk band structure. This is shown in Fig.~\ref{fig:sf-Chern} (Right). There are typically two possibilities of splitting of the doubly degenerate bands. The doubly degenerate bands can split along the rotation axes or perpendicular to the rotation axes. The two terms proportional to $\Gamma_{34}$ and $\Gamma_12$ split along the axes. We are currently unaware of bulk measurements  and will play an important role in clarifying the nature of the magnetic phase.  add parameters and y axis labels. }

%%%%%%%%%%%%%%%%%%%%%%%%%%%%%%%%%%%%

%%%%%%%%%%%%%%%%%%%%%%%
\twocolumngrid
\bibliography{CubicDSM}
 \end{document}